\documentclass[twocolumn]{aastex631}

\usepackage{xspace}

\newcommand{\Sec}[1]{{\protect\hyperref[sec:#1]{Sec.~\ref*{sec:#1}}}}
\newcommand{\Secs}[2]{{\protect\hyperref[sec:#1]{Secs.~\ref*{sec:#1}}~and~\ref{sec:#2}}}
\newcommand{\Fig}[1]{{\protect\hyperref[fig:#1]{Fig.~\ref*{fig:#1}}}}

\newcommand{\Equ}[1]{{\protect\hyperref[equ:#1]{Eq.~\ref*{equ:#1}}}}

\newcommand{\Tab}[1]{{\protect\hyperref[tab:#1]{Table~\ref*{tab:#1}}}}

\newcommand{\App}[1]{{\protect\hyperref[app:#1]{Appendix~\ref*{app:#1}}}}

\newcommand{\HI}{H{\footnotesize I}\xspace}

\newcommand{\Ha}{H\ensuremath{\alpha}\xspace}

\newcommand{\wunits}[2]{\ensuremath{#1\,\text{#2}}}
\newcommand{\magss}{{mag{\,}arcsec\ensuremath{^{-2}}}\xspace}

\newcommand{\probesdirty}{4208\xspace}
\newcommand{\probesnedmatched}{3293\xspace}
\newcommand{\probesnedunique}{3163\xspace}
\newcommand{\probesphotometric}{1677\xspace}
\newcommand{\probesrcextentre}{2\xspace}
\newcommand{\probesrcextentriso}{1\xspace}

\defcitealias{Courteau1997}{C97}

\received{15 May, 2022}
\revised{13 July, 2022}
\accepted{21 July, 2022}

\shorttitle{The PROBES Compendium}
\shortauthors{Stone et al.}

\begin{document}

\title{PROBES-I: A Compendium of Deep Rotation Curves and Matched multiband Photometry\footnote{Released on \today}}

\correspondingauthor{Connor Stone}
\email{connor.stone@queensu.ca}

\author[0000-0002-9086-6398]{Connor Stone}
\affiliation{Department of Physics, Engineering Physics and Astronomy,
  Queen{'}s University,
  Kingston, ON K7L 3N6, Canada}

\author[0000-0002-8597-6277]{St{\'e}phane Courteau}
\affiliation{Department of Physics, Engineering Physics and Astronomy,
  Queen{'}s University,
  Kingston, ON K7L 3N6, Canada}
  
\author[0000-0002-3929-9316]{Nikhil Arora}
\affiliation{Department of Physics, Engineering Physics and Astronomy,
  Queen{'}s University,
  Kingston, ON K7L 3N6, Canada}
  
\author[0000-0003-3506-0858]{Matthew Frosst}
\affiliation{Department of Physics, Engineering Physics and Astronomy,
  Queen{'}s University,
  Kingston, ON K7L 3N6, Canada}
\affiliation{ICRAR, M468, 
  University of Western Australia, 
  Crawley, WA 6009, Australia}

\author[0000-0002-4939-734X]{Thomas H. Jarrett}
\affiliation{Astronomy Department, University of Cape Town, 
  Rondebosch 7701, South Africa}

\begin{abstract}

We present the Photometry and Rotation Curve Observations from Extragalactic Surveys (PROBES) compendium of extended rotation curves for \probesnedunique late-type spirals, with matching homogeneous multiband photometry for \probesphotometric of them. 
PROBES rotation curves originally extracted from H$\alpha$ long-slit spectra and aperture synthesis HI (21cm) velocity maps typically extend out to a median {\probesrcextentre}R$_\textrm{e}$ (or {\probesrcextentriso}R$_{23.5, r}$).
Our uniform photometry takes advantage of GALEX, DESI-LIS, and WISE images and the software AutoProf to yield multiband azimuthally averaged surface brightness profiles that achieve depths greater than 25 \magss ({\it FUV, NUV}), 27 \magss ({\it g, r}), and 26 \magss ({\it z, W1, W2}).
With its library of spatially resolved profiles and an extensive table of structural parameters, the versatile PROBES data set will benefit studies of galaxy structure and formation. 
\end{abstract}


\section{Introduction}\label{sec:intro}

Galaxies are some of the best laboratories for studying fundamental questions about cosmology, dark matter, black holes, and more.
They are also complex objects with detailed physics occurring on scales spanning many orders of magnitude, and evolving in diverse environments, making the exercise of identifying the drivers of galaxy structure and evolution especially challenging. 
Teasing out and modelling evolutionary and transformative processes such as star formation~\citep{Salpeter1955,Tojeiro2009,Conroy2013}, merging~\citep{Toomre1972,Naab2014}, dust attenuation~\citep{Holmberg1946,Burstein1984,Tully1998,Stone2021a}, and angular momentum transport~\citep{Lin1987,Obreschkow2014} has deepened our understanding of the development of structure in the universe, while opening many new questions.

Chief among those, the elusive dark matter cannot be directly observed and must be inferred~\citep{Faber1979,Bertone2018,Wechsler2018}, despite its considerable effect on the evolution of galaxies.
The mismatch between the mass distribution through spatially resolved kinematic and photometric profiles is certainly one of the strongest evidence of dark matter in galaxies ~\citep{Bosma1978,Faber1979,vanAlbada1986,Courteau2014}.
While galaxy photometry is rather easily acquired, with large deep multi-wavelength imaging programs covering most of the sky~\citep{York2000,Skrutskie2006,Lawrence2007,Lang2014,Bianchi2017,Dey2019,Ivezic2019},
kinematics are more challenging given the targeted approach and time-consuming integrations, with most studies limited to a few hundred systems ~\citep{Rubin1980,Mathewson1992,Courteau1997,Sofue2001}.
Some recent programs have yielded large samples of galaxy kinematics with spatially resolved integral-field spectroscopy, though rarely reaching depths beyond $\sim$R$_{\rm e}$~\citep[SAMI:][]{Croom2021}, $\sim$1.5R$_{\rm e}$~\citep[MaNGA:][]{Bundy2015}, or $\sim R_{23.5}$~\citep[CALIFA:][]{Sanchez2016}. 
These efforts are nonetheless commendable and the future mapping of gravitational potentials in galaxies will require deep spatially resolved kinematics for tens of thousands of galaxies to achieve statistical power.

In this study, we take a step in this direction by providing one of the largest catalogues of galaxy light profiles and major-axis rotation curves (hereafter RCs) in the form of the ``Photometry and Rotation Curves Observations from Extragalactic Surveys (PROBES)" compendium.
PROBES combines some of the largest existing surveys of spatially resolved major-axis rotation curves (hereafter RCs) into a single homogenized compilation of \probesnedunique spiral galaxies. 
We have taken advantage of the GALEX Survey~\citep{Bianchi2017}, DESI Legacy Survey~\citep{Dey2019}, and the unWISE survey~\citep{Lang2014} to compliment the kinematic information with homogeneous multiband data from the UV to the NIR in the {\it FUV, NUV, g, r, z, W1}, and {\it W2} bands. 
With this publication, we make the PROBES compilation readily available to the community.

The PROBES combination of deep photometry and kinematic information for a large number of spiral galaxies is ideally suited for studying the interplay between baryons and dark matter.
The PROBES sample has already been included in various studies during its development.
\citet{Stone2019} examined the Radial Acceleration Relation of PROBES galaxies, favouring a $\Lambda$CDM interpretation over Modified Newtonian Dynamics~\citep{Lelli2017}.  
PROBES galaxies also allowed \citet{Stone2021a} to find that scatter had been systematically underestimated in previous analyses of galaxy structural scaling relations.
PROBES scaling relations have been used as a benchmark for the analysis of MaNGA data \citet{Arora2021} and to characterize the diversity of spiral galaxies \citet{Frosst2022}.

This paper is divided as follows.
\Sec{probessources} introduces the surveys amalgamated into the PROBES compendium, emphasizing the respective selection criteria.
A variety of structural parameters extracted from the light profiles and RCs for these galaxies are then presented in \Sec{structuralparameters}. 
The data tables and readme files provided with the PROBES compendium are described in \Sec{datatables}.
We conclude in \Sec{conclusion} by reiterating the usefulness of these data for studies of galaxy structure and formation, stellar populations, galaxies as cosmological tracers, and more. 

\section{PROBES Data Sources}\label{sec:probessources}

\subsection{Sample Selection}\label{sec:sampleselection}

The PROBES compendium is a combination of seven previously published RC surveys with new {\it FUV, NUV, g, r, z, W1}, and {\it W2} matched photometry presented here, combined and homogenized for easier usability.
Brief descriptions of the seven surveys, with an emphasis on sample selection\footnote{Replicating some information from \citet{Stone2019} for completeness.}, are presented below.
Most of the original surveys provided their own photometry in addition to the RCs; however, for homogeneity, we have extracted and standardized our own photometry from the DESI-LIS~(\Sec{desiphotometry}), unWISE~(\Sec{unwisephotometry}), and GALEX~(\Sec{galexphotometry}) surveys instead.
Of the \probesphotometric PROBES galaxies with {\it r}-band photometry, $>99\%$ have {\it g, z, W1, } and {\it W2} bands while $~80\%$ have {\it FUV, NUV} bands.
While DESI-LIS and unWISE images enable detailed studies of their respective spatially resolved light profiles, the noisier GALEX images will only be used to compute global quantities such as total luminosities and to complement spectral energy distributions (SEDs).  

\subsubsection{Mathewson 1992}

\citet{Mathewson1992} published \Ha RCs for 827 spiral galaxies found largely in the southern hemisphere. 
Galaxies were selected primarily from the ESO-Uppsala catalog \citep{Lauberts1982,Lauberts1998} with morphological types shown in \Fig{hubbletypes}, diameters greater than \wunits{1.7}{arcmin}, 
heliocentric recessional velocities below \wunits{7,000}{km\,s$^{-1}$}, moderate inclinations, and Galactic latitude $|b| > \wunits{11}{deg}$. 
Some galaxies were taken from other surveys~\citep{Mathewson1992}.
The velocity uncertainties for these RCs were not reported, instead we use the residuals from our model fits (see \Sec{evalvelocity}) to estimate reasonable upper bound uncertainties.

\subsubsection{Mathewson 1996}

\citet{Mathewson1996} expanded the \citet{Mathewson1992} sample to 2017 spiral galaxies with \Ha RCs primarily located in the southern hemisphere.
The morphological distribution is shown in \Fig{hubbletypes}.
However, not all objects in \citet{Mathewson1992} were included in \citet{Mathewson1996}.
As in \citet{Mathewson1992}, velocity uncertainties for RCs were not provided and a model fit was used to determine upper bound uncertainties (see \Sec{evalvelocity}). 
The sampling criteria were similar to those of \citet{Mathewson1992}, though with higher heliocentric recessional velocities in the range \wunits{4,000–14,000}{km\,s$^{-1}$} and apparent diameters between \wunits{1.0-1.6}{arcmin}. 
Once again, a small number of galaxies were taken from other surveys, such as the Catalogue of Galaxies and Clusters of Galaxies~\citep{Zwicky1961,Zwicky1995}.

\subsubsection{Courteau 1997}

The \citet[][hereafter \citetalias{Courteau1997}]{Courteau1997} sample is a survey of 353 late-type galaxies with \Ha RCs and morphological types shown in \Fig{hubbletypes}. 
These were collected largely for cosmic flow studies for which systematic and random velocity uncertainties were of great interest \citep{Courteau1993}. 
Thus, many galaxies have multiple measurements, with some having as many as four repeat RCs, and well characterized velocity uncertainties. 
For the PROBES compendium, all multiple RC measurements, were stacked as described in \Sec{rotationcurves}.
The sample was selected from the Uppsala Catalog of Galaxies~\citep{Nilson1973,Lauberts1998} and the catalog of cluster galaxies from~\citep{Bothun1985}. 
\citetalias{Courteau1997} galaxies were selected to have Hubble types Sb–Sc, Zwicky magnitude $m_B\leq15.5$~\citep[see][]{Giovanelli1984}, blue Galactic extinction less than \wunits{0.5}{mag} (based on \citealt{Burstein1984}), inclinations between \wunits{55-75}{deg}, blue major axes less than \wunits{4}{arcmin}, no interactions or mergers, and no bright foreground stars \citep{Courteau1996}.

\subsubsection{SCII 1999}

The all-sky survey of \citet[][hereafter SCII]{Dale1999} offers \Ha RCs for 602 galaxies in 52 Abell clusters up to heliocentric recessional velocities of \wunits{25,000}{km\,s$^{-1}$}. 
Galaxies were selected from the Abell Rich Cluster Catalog \citep{Abell1989}, favoring those with redshift information available at the time.
Morphologies for the SCII galaxies are shown in \Fig{hubbletypes}.
These galaxies add a significant number of late-type cluster member galaxies to the PROBES compendium, allowing for environmental studies among others.
We use the velocity errors as reported in the original data.

\subsubsection{Shellflow 2000}

The Shellflow survey \citep{Courteau2000} of 186 spiral galaxies was designed to study an all-sky shell in redshift space to measure a cosmological bulk flow of galaxies with high precision. 
The Shellflow sample geometry meant that a large fraction of the galaxies could be observed from both northern and southern hemisphere observatories, namely the Kitt Peak National Observatory and Cerro Tololo Inter-American Observatory, thus mitigating calibration errors from different instrumentation. 
The Shellflow sample was selected from the Optical Redshift Survey of \citet{Santiago1995}. 
Galaxies were chosen to be noninteracting and of morphological types indicated in \Fig{hubbletypes}, with heliocentric recessional velocities between \wunits{4,500 - 7,000}{km\,s$^{-1}$}, inclinations between \wunits{45 - 78}{deg}, AB extinctions less than \wunits{0.3}{mag} (as determined by \citealt{Burstein1982}), and no bright overlapping foreground stars or tidal disturbances.
We use the velocity errors as reported in the original data.

\subsubsection{SPARC 2016}\label{sec:sparc}

The Spitzer Photometry and Accurate Rotation Curves (SPARC) sample compiled by \citet{Lelli2016} is an amalgamation of over 50 smaller samples totaling 163 galaxies with a mix of \Ha and \HI RCs.
Approximately one-third of the SPARC galaxies have hybrid \HI and \Ha RCs to combine the higher spatial resolution provided by \Ha with the extensive radial extent of synthesis \HI radio maps, where available. 
The distance estimates for SPARC galaxies rely on a number of methods, including Hubble flow, tip of the red giant branch, Cepheids, Ursa Major cluster distance, and supernovae. 
All distances and their uncertainties are reported in the SPARC survey, and we use them directly in the PROBES compendium. 
The sample was carefully chosen to have a range of morphology, luminosity, and surface brightness (SB) from the limited selection of galaxies with \HI measurements \citep{Lelli2016}. 
We use the velocity errors as reported in the original data.

\subsubsection{SHIVir 2017}

The Spectroscopy and H-band Imaging of Virgo Cluster Galaxies (SHIVir) survey presented in \citet{Ouellette2017} and \citet{Ouellette2022} is a dedicated survey of galaxies of all morphological types in the Virgo Cluster (see \Fig{hubbletypes}).
SHIVir's original purpose was to examine the impact of a cluster environment on galaxy properties using the nearest cluster in the sky: the Virgo Cluster. 
We focus on the subset of 44 SHIVir spiral galaxies with \Ha RCs from long-slit spectra taken at various observatories \citep{Ouellette2017}.
We use the velocity errors as reported in the original data.
Distances to the SHIVir galaxies were estimated from SB fluctuations where available \citep{Jerjen2004,Blakeslee2009}; otherwise, a standard cluster distance of \wunits{16.5}{Mpc} is assumed \citep{Mei2007}. 
Uncertainties are reported where distance measurements are available; for the cluster distance, a conservative value of \wunits{3}{Mpc} was used.

The SHIVir galaxies are a subset of the full Virgo Cluster Catalog (VCC), which is volume complete in a spatial subset of the Virgo Cluster to an absolute magnitude of $M_B \leq -13$, along with several fainter galaxies to ensure a broad morphology coverage \citep{Binggeli1985,McDonald2011}.

\subsection{NED Information and Quality Criteria}\label{sec:neddata}

\begin{figure}
    \centering
    \includegraphics[width=1.05\columnwidth]{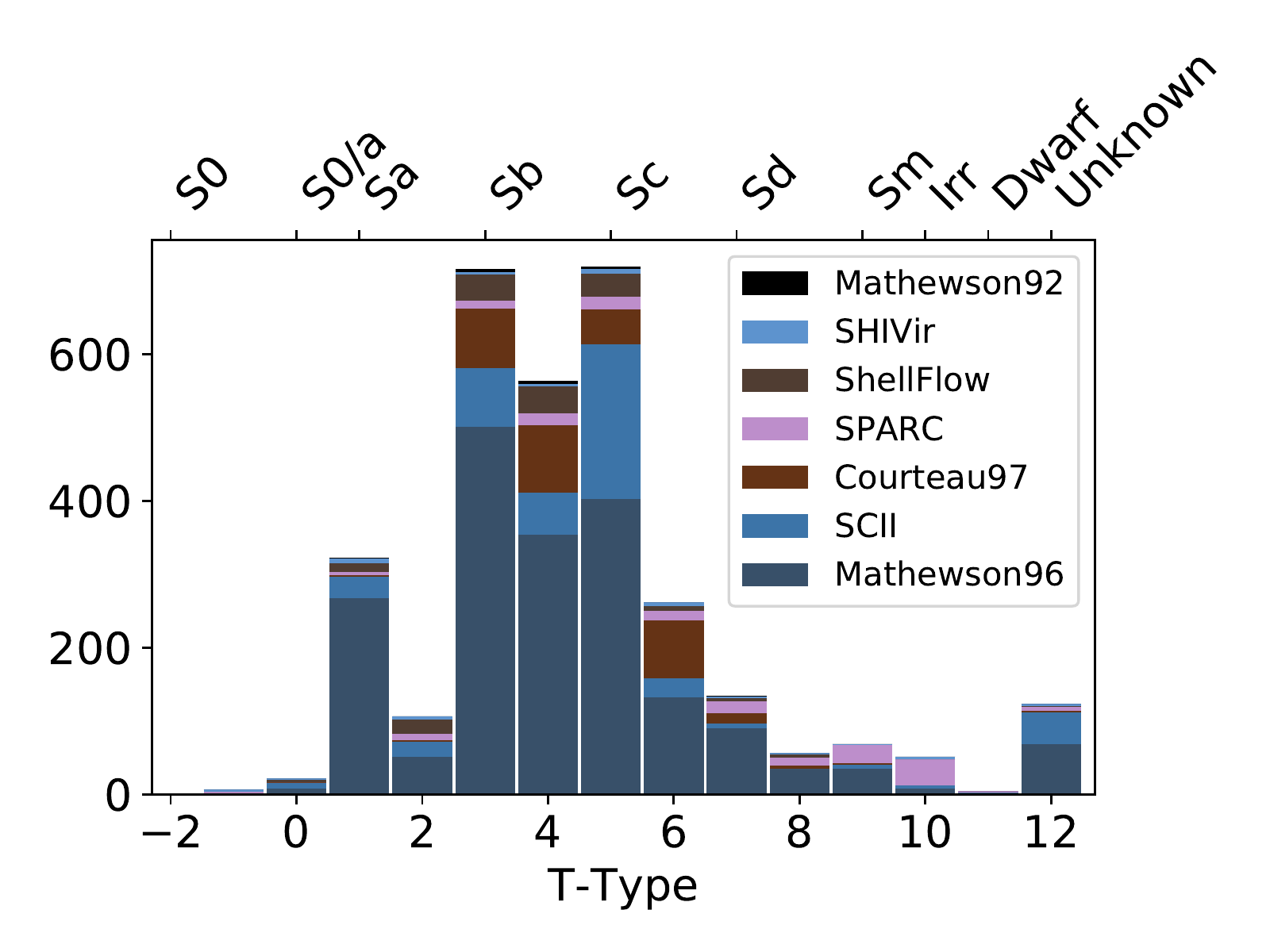}
    \caption{Distribution of morphological T-types for the PROBES photometric sample. 
    The legend gives the surveys in order of total number of galaxies.
    The distribution is peaked for Sb and Sc types, with significant contributions from the Mathewson and SCII samples. 
    A small number of galaxies have no known morphology in NED.}
    \label{fig:hubbletypes}
\end{figure}

The cross-referencing of all galaxies from the seven surveys above with NED\footnote{The NASA/IPAC Extragalactic Database (NED) is operated by the Jet Propulsion Laboratory, California Institute of Technology, under contract with the National Aeronautics and Space Administration.} yielded \probesdirty galaxies with names, coordinates, and redshift.  
Most galaxies had morphological information, as presented in \Fig{hubbletypes} where T-type 12 indicates no available data (3.3\% of the PROBES sample).
Objects were matched by name and confirmed using redshift information in NED and velocity information from the observed RC.
In some instances, only RA and DEC coordinates were usable; naming conventions for some older surveys are no longer recognized. 
Again these were confirmed using redshift information (which is available for all PROBES galaxies).
If neither a name nor coordinates could be confidently confirmed, the galaxy was discarded from the survey; 
this yielded a total sample of \probesnedmatched galaxies.

Furthermore, many galaxies were duplicated between surveys.
In such instances, we stacked their RCs as described in \Sec{rotationcurves}. 
A label, called ``RC\_survey'', in the main table indicates that multiple surveys contributed to the final RC.
The exception is the \citet{Mathewson1992} and \citet{Mathewson1996} samples for which the \cite{Mathewson1996} RCs supersede the other; most of the repeated galaxies between the two surveys are indeed from the same observations.
Thus we are left with a PROBES sample of \probesnedunique unique galaxies with RCs and matched NED metadata.

\subsection{Distances}\label{sec:distances}

The PROBES compendium distances come from a number of sources including Hubble flow, surface brightness fluctuations, and tip of the red giant branch.
For Hubble flow distances, we have used the CosmicFlows-3 calculator from \citet{Kourkchi2020a}\footnote{see the following link: \url{https://edd.ifa.hawaii.edu/CF3calculator/}}.
Beyond \wunits{200}{Mpc}, we have assumed a Hubble-Lema{\^i}tre constant of $H_0 =$~\wunits{73}{km s$^{-1}$ Mpc$^{-1}$}~\citep{Riess2021}. 
Heliocentric redshifts were taken from NED and corrected to cosmic microwave background (CMB) redshifts using a standard \citet{Fixsen1996} apex velocity.

\subsection{Rotation Curves}\label{sec:rotationcurves}

The PROBES compendium RCs come mainly from \Ha long-slit (major axis) spectra, though a few \HI RCs were also available.
RC measurements are obtained from an observed wavelength, $\lambda_o$ (e.g. from a long slit spectrograph), and converted to a relative velocity, $v$, using the formula:

\begin{equation}
    \frac{\lambda_o}{\lambda_s} = \sqrt{\frac{1+v/c}{1-v/c}} \approx 1 + v/c,
\end{equation}

where $\lambda_s$ is the source wavelength.
In the local universe, where PROBES RCs are measured, this approximation is accurate enough for most purposes.
At cosmological distances, the recession velocity is a poorly defined quantity since the observed wavelength shift is caused by a combination of relativistic Doppler shift and stretching of spacetime. 
In such instances, the well defined quantity redshift, $z$, can then be better related to velocities for the construction of an RC~\citep{Bunn2009}.
In the PROBES regime, the contribution due to stretching of spacetime can simply be treated as a component of the systemic velocity.

\begin{figure}
    \centering
    \includegraphics[width=\columnwidth]{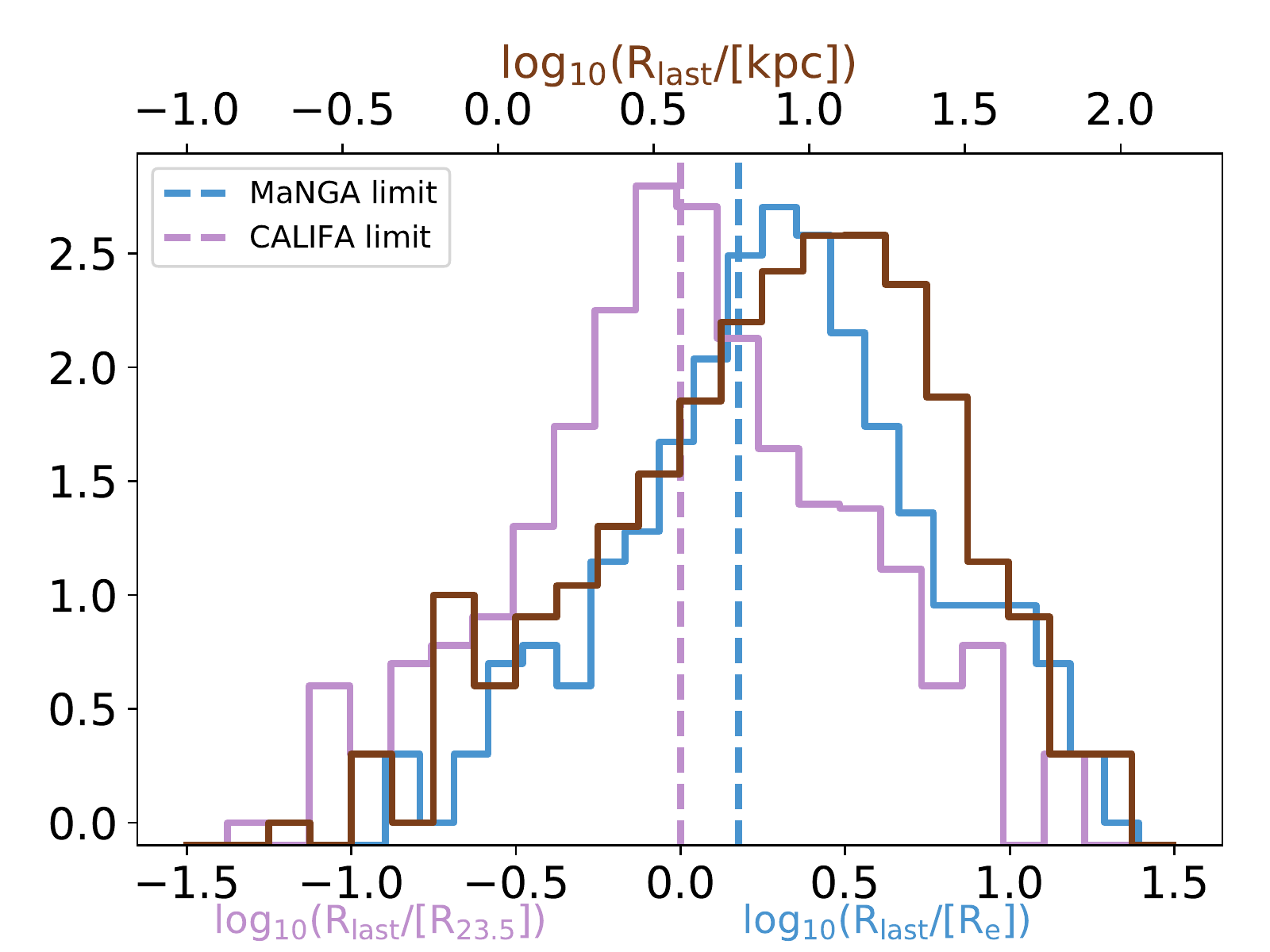}
    \caption{Distribution of RC extent in the PROBES data set. 
    RC extent is the radial value of the last data point in each RC. 
    The distribution is given (top axis) in physical size (kpc),  or normalized (lower axis) by a fiducial radius ($R_{\rm e}$ and R$_{23.5}$). 
    Most RC profiles extend beyond $2R_{e}$, or beyond R$_{23.5}$.
    The nominal extents of MaNGA and CALIFA galaxies are shown for reference.}
    \label{fig:rlast}
\end{figure}

\Fig{rlast} shows the spatial extent distribution of RCs that have corresponding photometry.
The brown distribution normalizes $R_{\rm last}$ in kpc; normalizations by effective radius, $R_{\rm e}$, and isophotal radius, R$_{23.5}$, are also presented.
The median extent of the PROBES galaxies is {\probesrcextentre}$R_{\rm e}$ or {\probesrcextentriso}R$_{23.5}$.
This contrasts with previous IFU surveys such as CALIFA~\citep{Sanchez2016} and MaNGA~\citep{Bundy2015} which extend to R$_{23.5}$ and $1.5R_{\rm e}$, respectively. 
With such deep measurements, the extended PROBES RCs can efficiently probe the flatter region of a galaxy disk beyond the transition from baryon- to dark matter-dominated systems.

Our RCs were all fitted with a \textit{tanh} model and the multiparameter model of  \citetalias{Courteau1997}.
The \textit{tanh} model matches the simple rise and flattening idealization of each RC.
The second, \citetalias{Courteau1997} multiparameter model, here referred to as the ``\citetalias{Courteau1997} model'', provides a more complete model of the rich variety of RC shapes \citep{Oman2015,Frosst2022}.
Further details about these fits are given in \Sec{evalvelocity}. 

\begin{figure*}
    \centering
    \includegraphics[width=0.8\textwidth]{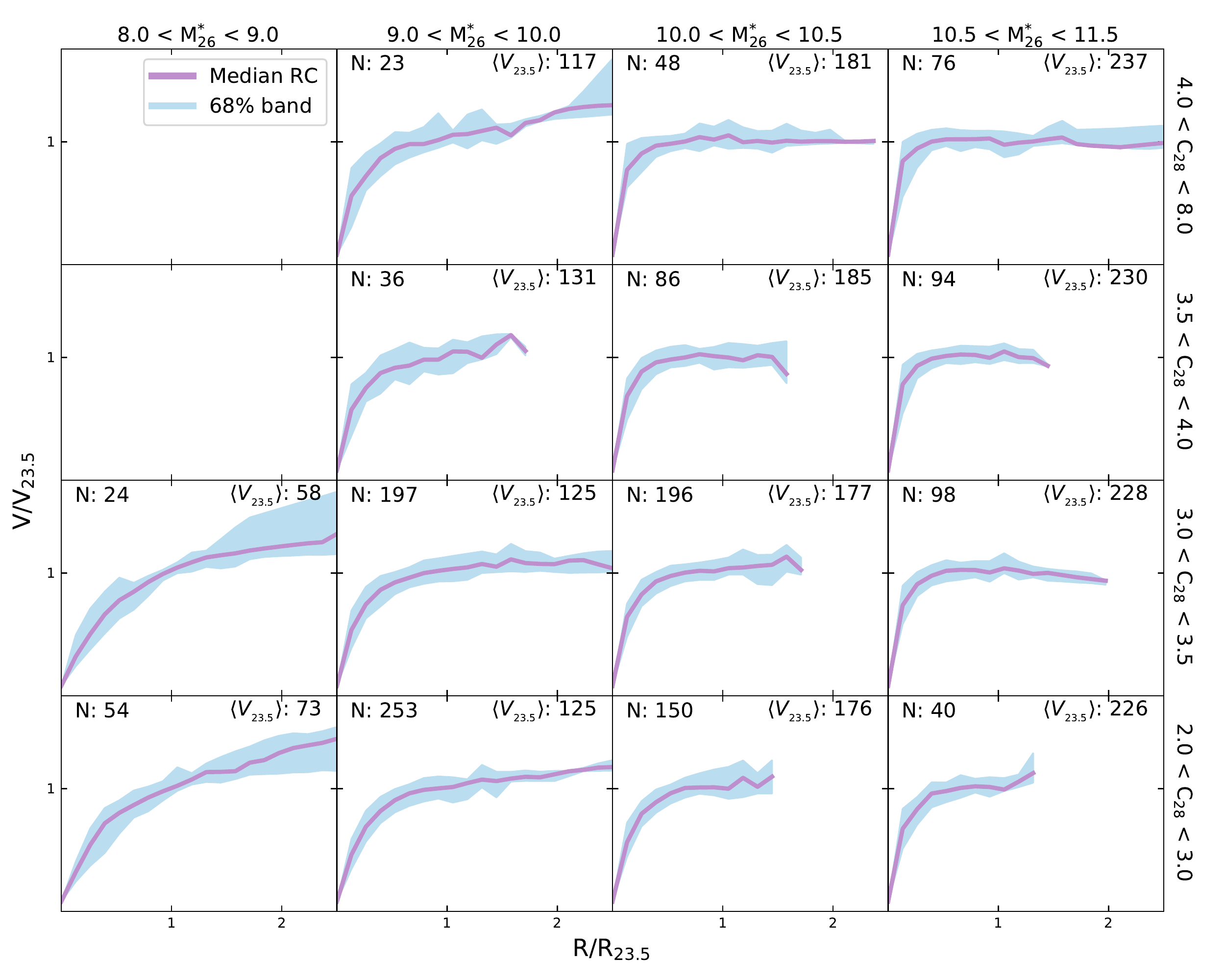}
    \caption{Median RCs in the PROBES compendium in stellar mass and concentration bins specified by the top and right labels.
    The radii and velocities were normalized (before stacking) by the {\it r}-band isophotal radius $R_{23.5}$ and velocities measured at $R_{23.5}$, $V_{23.5}$, respectively.
    The number of galaxies included in each bin and mean $V_{23.5}$ of the median stack are shown as text inset in the upper left and right sides of each panel.
    The radial scale is limited to $2.5R/R_{23.5}$ to emphasize the range of shapes in the inner RCs, though many RCs extend beyond this point.}
    \label{fig:rcgrid}
\end{figure*}

\Fig{rcgrid} shows median RC shapes in the PROBES compilation in bins of stellar mass (columns) or concentration (rows).
Each RC is linearly interpolated to a standard set of radii, and the median is taken for all galaxies which have measurements at those radii.
For this figure, stacked RCs were truncated if less than 3 measured profiles contributed to the median profile.
The profile radii and velocities were normalized by $R_{23.5}$ and $V_{23.5}$, respectively (see \Sec{structuralparameters}).
This enables comparisons on the basis of changing shapes alone.
Each panel corresponds to a range of concentration and stellar mass values (see \Sec{structuralparameters}) and a clear trend in RC shape can be seen.
For higher stellar mass (concentration) at a given concentration (stellar mass), the RC rises faster near the center~\citep{Rubin1985,Madore1987,Persic1996,Sofue2001}.
In all but the lowest stellar mass bin, the RCs flatten at or before approximately $R_{23.5}$; indeed, \Fig{rlast} shows that the majority of PROBES RCs extend to the flattened region.

\begin{figure}
    \centering
    \includegraphics[width=\columnwidth]{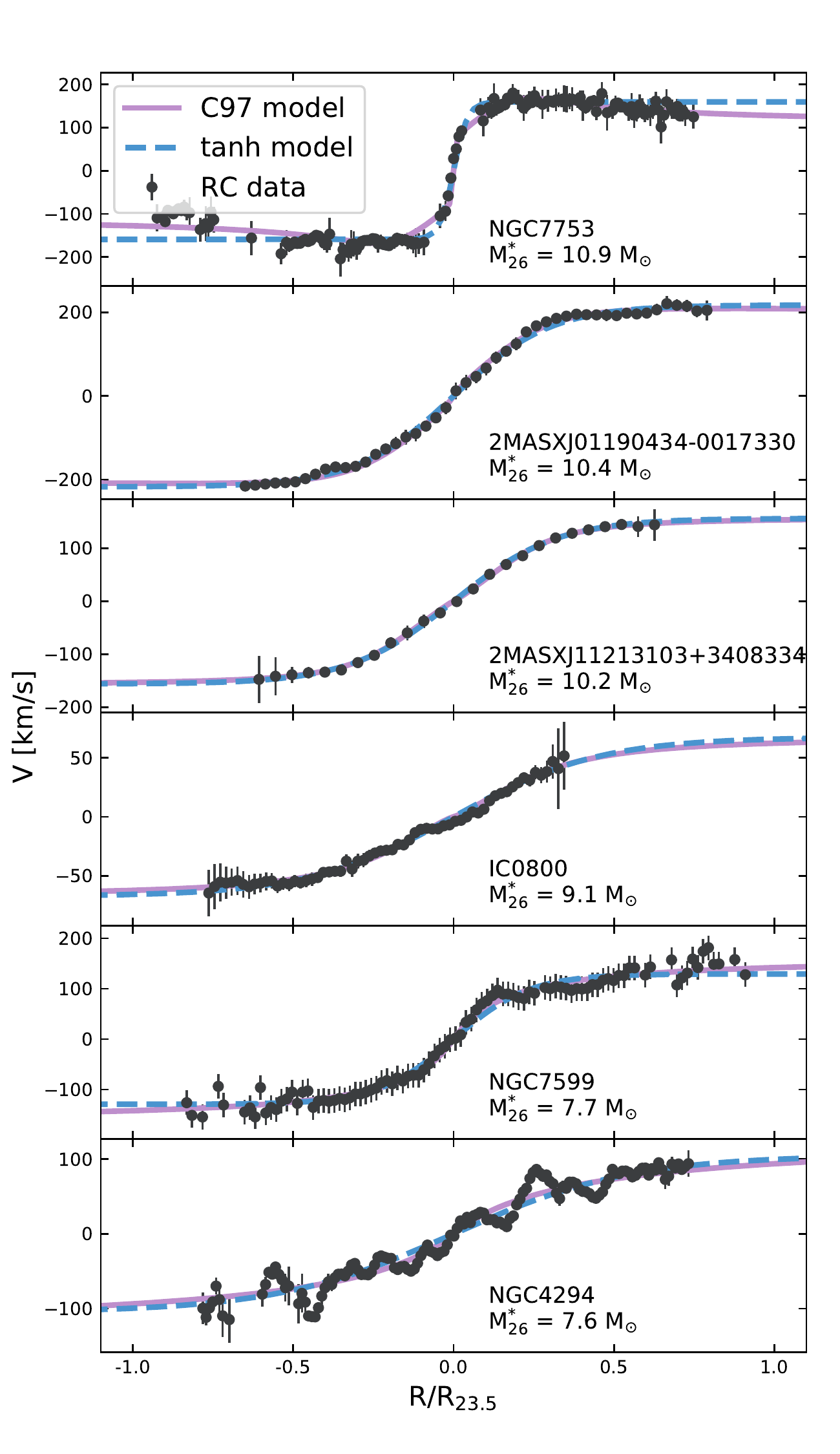}
    \caption{Example RCs from PROBES over a broad range of stellar masses, M$^*_{26}$, with \citetalias{Courteau1997} and \textit{tanh} model fits. 
    Data points are taken directly from PROBES profiles and the errors are scaled by a factor of 2 for visibility.  
    PROBES RCs display a great range of shapes and sampling densities~\citep[see also][]{Frosst2022}.}
    \label{fig:examplercs}
\end{figure}

\Fig{examplercs} gives example PROBES RCs for galaxies with a range of stellar masses. 
As in \Fig{rcgrid}, a range of RC shapes can be seen from fast rising to very gradual when normalized by $R_{\rm 23.5}$.
The highest stellar mass example, \emph{NGC7753}, shows a velocity peak at about $0.2R_{23.5}$.
The PROBES RCs are densely sampled in most cases allowing for more advanced kinematic analysis.
In all cases the \citetalias{Courteau1997} model fits and the \textit{tanh} fits agree well within the bounds of the data, though each yield slightly different extrapolations at large radii.
Velocity bumps and wiggles, indicative of noncircular motions and other complex kinematic structure, especially near the galaxy center, are not modelled by the  \citetalias{Courteau1997} or \textit{tanh} models. 

\subsection{DESI Legacy Imaging Survey Photometry}\label{sec:desiphotometry}

\begin{figure*}
    \centering
    \includegraphics[width=\textwidth]{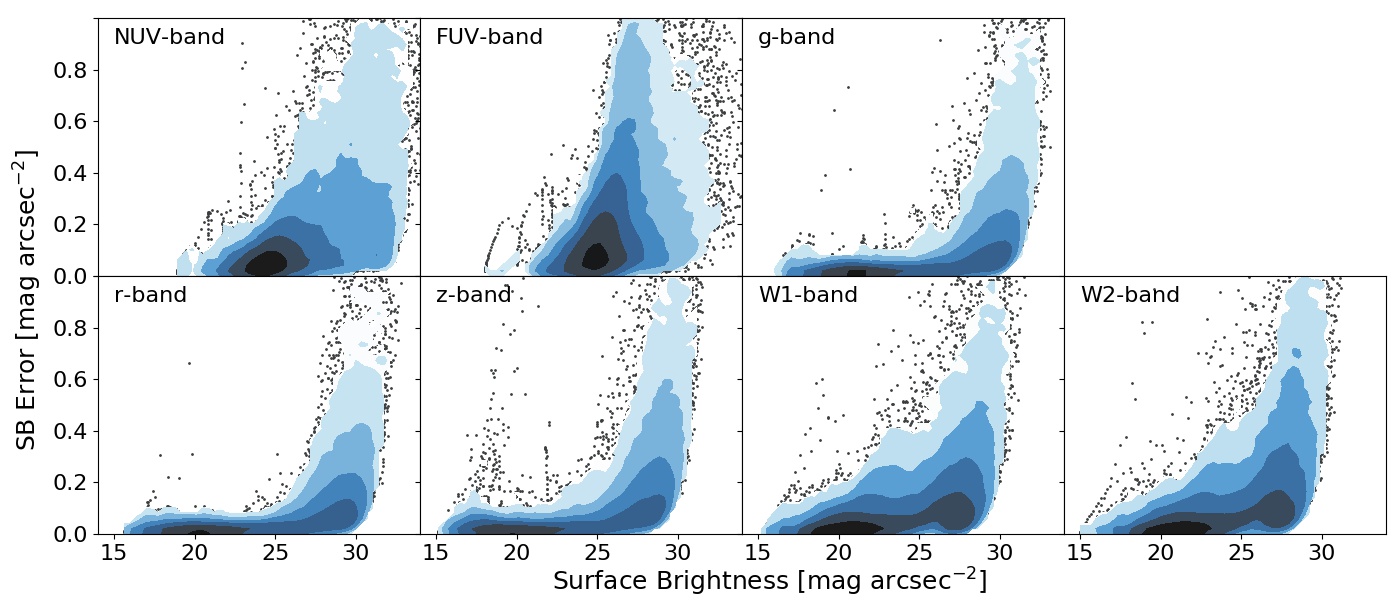}
    \caption{Comparison of SB errors for the multiband GALEX, DESI-LIS, and unWISE photometry as a function of SB depth. 
    Scatter points were taken from every profile in the PROBES photometry sample, while contours were used in high density regions. 
    Most SB profiles reach at least \wunits{26}{\magss} before errors start rising; some profiles reach \wunits{30}{\magss} with small errors.
    This deep multiband photometry critically maximizes the scientific potential of the PROBES photometric dataset.}
    \label{fig:photometryerrors}
\end{figure*}

The DESI Legacy Imaging Survey provides {\it g, r, z} photometry for a large \wunits{14,000}{deg$^2$} area of the sky~\citep{Dey2019}.
With some of the PROBES galaxies lying outside the DESI-LIS footprint, we have constructed a ``PROBES photometry'' sample of \probesphotometric galaxies for which matching DESI-LIS photometry can be extracted.
The automated nonparametric surface photometry package AutoProf~\citep{Stone2021b} was then applied to all available DESI-LIS {\it r}-band images, achieving SB thresholds of $\sim$28 {\it r}-\magss (see \Fig{photometryerrors}, all photometry uses the AB magnitude system). 
AutoProf performs nonparametric isophotal ellipse fitting by minimizing the power in the second Fourier mode along each ellipse as a function of PA and ellipticity.
Azimuthally averaged profiles as well as cardinal wedges (along the four major/minor axes from the center of the galaxy) were extracted for all PROBES photometric galaxies. 
These wedges offer rich information about the symmetry of galaxies and detailed nonaxisymmetric structure, where present.
A unique PSF is determined for each observation, however DESI-LIS images typically have a FWHM resolution of $\sim$\wunits{1}{arcsec}.

The fitted isophotal solution for the {\it r}-band image of a galaxy serves as a reference for the matching {\it g} and {\it z} band images for the same galaxy. 
The ``forced photometry'' procedure with AutoProf ensures that all the flux values are calculated along the same isophotes for a given galaxy, a requirement for meaningful colour (and therefore stellar mass) measurements.
The AutoProf isophotal fitting is entirely automated though it can fail under certain conditions~\citep[e.g. overlapping systems, see][]{Stone2021b}.
These conditions are flagged by AutoProf (also automatically);
all flags are reported in the main PROBES table (see \Sec{datatables}).

To characterize the depth of the DESI-LIS/AutoProf photometry, we present in \Fig{photometryerrors} SB errors versus the corresponding SB depth of every isophote in the sample. 
A clear rise in errors is detected beyond $\sim$\wunits{27}{\magss} ({\it g}-, {\it r}-band) or \wunits{26}{\magss} ({\it z}-band).
Still, some isophotes reach \wunits{30}{\magss} with small errors.
These data are thus deep enough for a broad suite of structural analyses.
Note that the feature seen in the {\it z}-band subplot of \Fig{photometryerrors} with high brightness and high error is caused by bright galaxies which saturate the detectors.
Saturated pixels register in AutoProf as high errors at the transition from saturation to real measurements.
This feature is also seen, though to a lesser degree, in the {\it g} and {\it r} bands. 

\subsection{unWISE Photometry}\label{sec:unwisephotometry}

The unWISE survey provides all-sky {\it W1} and {\it W2} band photometry at a resolution of $\sim$\wunits{6}{arcsec}; fluxes are converted to the AB magnitude system for homogeneity.
The mid-infrared photometry available through the unWISE survey is sensitive to older, mass dominant, stellar populations resulting in robust stellar mass measurements \citep{Conroy2013, Courteau2014}.
Most PROBES galaxies are close enough that they can be resolved by unWISE photometry.
We have thus performed forced photometry (again using the {\it r}-band isophotal solutions as reference) to obtain SB profiles for the entire PROBES photometry sample.
Given the low resolution of the unWISE data, this addition mostly benefits total luminosity measurements, allowing for more robust modelling of the SED
(the same will be true for GALEX data; see \Sec{galexphotometry}).
Other structural parameters computed at large radii are also minimally affected by the low resolution.

As for our DESI-LIS data, \Fig{photometryerrors} shows the depth of all extracted unWISE SB profiles.
The unWISE SB profiles are shallower than the matching DESI-LIS profiles, however they exhibit the same general trend with somewhat higher errors before rising quickly at a limiting SB level of $\sim$27 {\it r}-mag arcsec$^{-2}$.
The {\it W1} and {\it W2} bands extend deep enough for robust measurements of total light, however the low spatial resolution limits the use of structural metrics at low radii.

We forgo the addition of NIR (JHK) bands as currently available surveys lack the desired depth~\citep{Skrutskie2006} or sky coverage~\citep{Lawrence2007}.

\subsection{GALEX Photometry}\label{sec:galexphotometry}

The GALEX survey provides nearly all-sky FUV and NUV ({\it FUV} and {\it NUV}, respectively) broadband photometry at a resolution of approximately 4.3 and \wunits{5.3}{arcsec}, respectively; this photometry is also converted to the AB magnitude system for homogeneity. 
Most PROBES galaxies are close and bright enough for GALEX forced photometry using AutoProf.
As with the {\it g, z, W1,} and {\it W2} bands, the {\it r}-band isophotal solutions were applied via forced photometry to the GALEX {\it FUV} and {\it NUV} images for uniform colours.
\Fig{photometryerrors} shows the GALEX SB measurement quality.
Similar to the unWISE data, the low GALEX resolution thwarts meaningful measurements of structural parameters in the UV; especially near the center of each galaxy. 
However, GALEX data provide useful SED constraints through bands especially sensitive to dust and recent star formation.

\section{Structural Parameters}\label{sec:structuralparameters}

In addition to the suite of RCs and SB profiles, the PROBES compendium offers a comprehensive array of galaxy structural parameters.
These include multiple definitions of size, luminosity, velocity, surface density, stellar mass, concentration, and so on.

This section presents a description of the PROBES parameters, how they are calculated and corrected.
Our tables include standard parameters evaluated at common radii ($R_{e}$, $R_{23.5}$, $R_{1}$, etc.) in the $r-$band.

\subsection{Corrections}\label{sec:paramcorrections}

Before we proceed, a discussion of the corrections to photometric and kinematic measurements is warranted. 
These corrections are applied to the RCs and SB profiles before computing any parameters.
Note that the PROBES database of RCs and SB profiles is presented uncorrected (raw),
the only exception being the RC data that are presented with the systemic velocity, $V_{sys}$, removed (\Sec{evalvelocity}).
The following discussion details how the corrections for our structural parameters were computed.

\subsubsection{Velocity Corrections}\label{sec:velocitycorrection}

Observed radial velocities, $V_o(R)$, are deprojected and corrected for redshift broadening according to: 

\begin{equation}
    V_c(R) = \frac{V_o(R) - V_{sys}}{\sin(i_{\rm last})(1+{\rm z}_{\rm helio})}
\end{equation}

where $V_{c}(R)$ is the corrected line-of-sight velocity, $V_{sys}$ is the fitted systemic (central) velocity from the \citetalias{Courteau1997} model (see \Sec{evalvelocity}), ${\rm z}_{\rm helio}$ is the heliocentric redshift of the galaxy, and $i_{\rm last}$ is the nominal inclination of the galaxy disk (the inclination of the outermost isophote).
The values for the systemic velocities were determined from the fitted RC model. 
Literature values for the heliocentric velocity of a galaxy are not used for $V_{sys}$ in order to avoid wavelength calibration differences.

\subsubsection{Surface Brightness Profile Truncation}\label{sec:profiletruncation}

\begin{figure}
    \centering
    \includegraphics[width=\columnwidth]{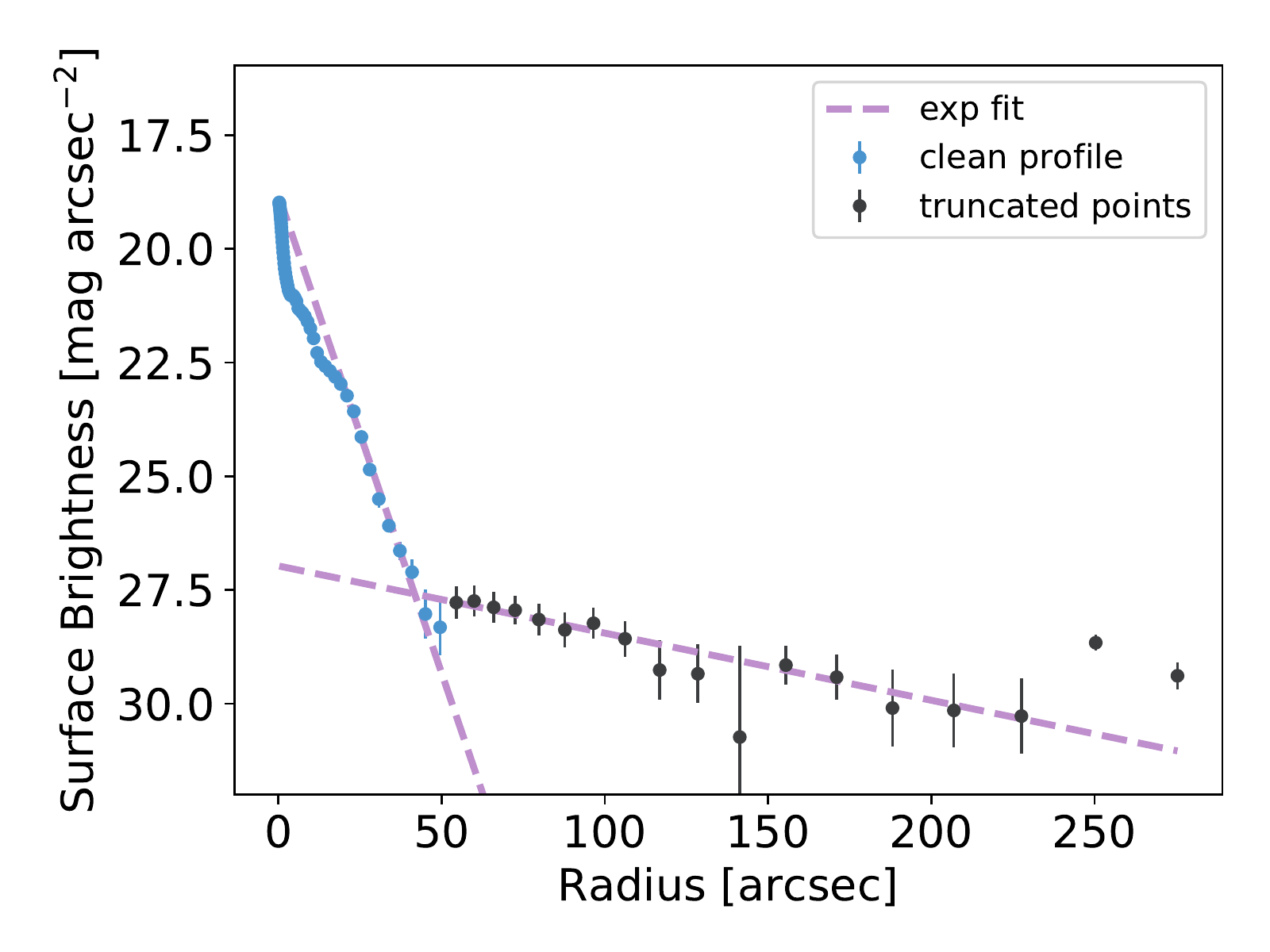}
    \caption{Example {\it r}-band profile truncated at the noise floor for ESO143-G028. 
    The purple dashed lines give the exponential disk fit to the outer profile (blue points) and the SB floor (grey points). 
    The data beyond the intersection of the exponential disk and the SB floor are ignored for further analysis. 
    Error bars are multiplied by 5 for visibility.} 
    \label{fig:truncatedprofile}
\end{figure}

To avoid aberrant isophotes from contaminating parameter estimations, SB profiles are truncated if the galaxy is no longer the dominant light source. 
Isophotes with an uncertainty exceeding \wunits{0.3}{\magss} for {\it g, r, z, W1, W2} bands, or \wunits{0.5}{\magss} for {\it FUV, NUV} bands, are first removed from the profile.
As seen in \Fig{photometryerrors}, errors rise quickly beyond $\sim$0.2~\magss.
We adopt a conservative \wunits{0.3}{\magss} threshold for our truncation level.
The SB errors are statistical in nature and do not account for spurious factors such as nearby objects, bright stars, and galactic cirri which can cause systematic fluctuations that are unrepresentative of the galaxy disk. 
To handle these cases, we fit an exponential disk + floor model to the {\it r}-band SB profile outskirts\footnote{The outskirts are initially defined as the point beyond which the SB profile is a factor of 100 (\wunits{5}{\magss}) dimmer than the central region.
This is intended to avoid bright extended bulges and other nonexponential central features.}.
The ``floor'' model is also an exponential with a shallow slope to represent the sky background plus interlopers around the galaxy.
Once fit, we truncate at the intersection between the exponential disk model and the floor value.
Points are reintroduced if they align closely ($\pm 1$ \magss) with the inner exponential fit.
This truncation radius in the {\it r}-band is then applied to all other bands.
\Fig{truncatedprofile} shows an example of a truncation which prevents spurious background data from entering further calculations.

\begin{figure}
    \centering
    \includegraphics[width=\columnwidth]{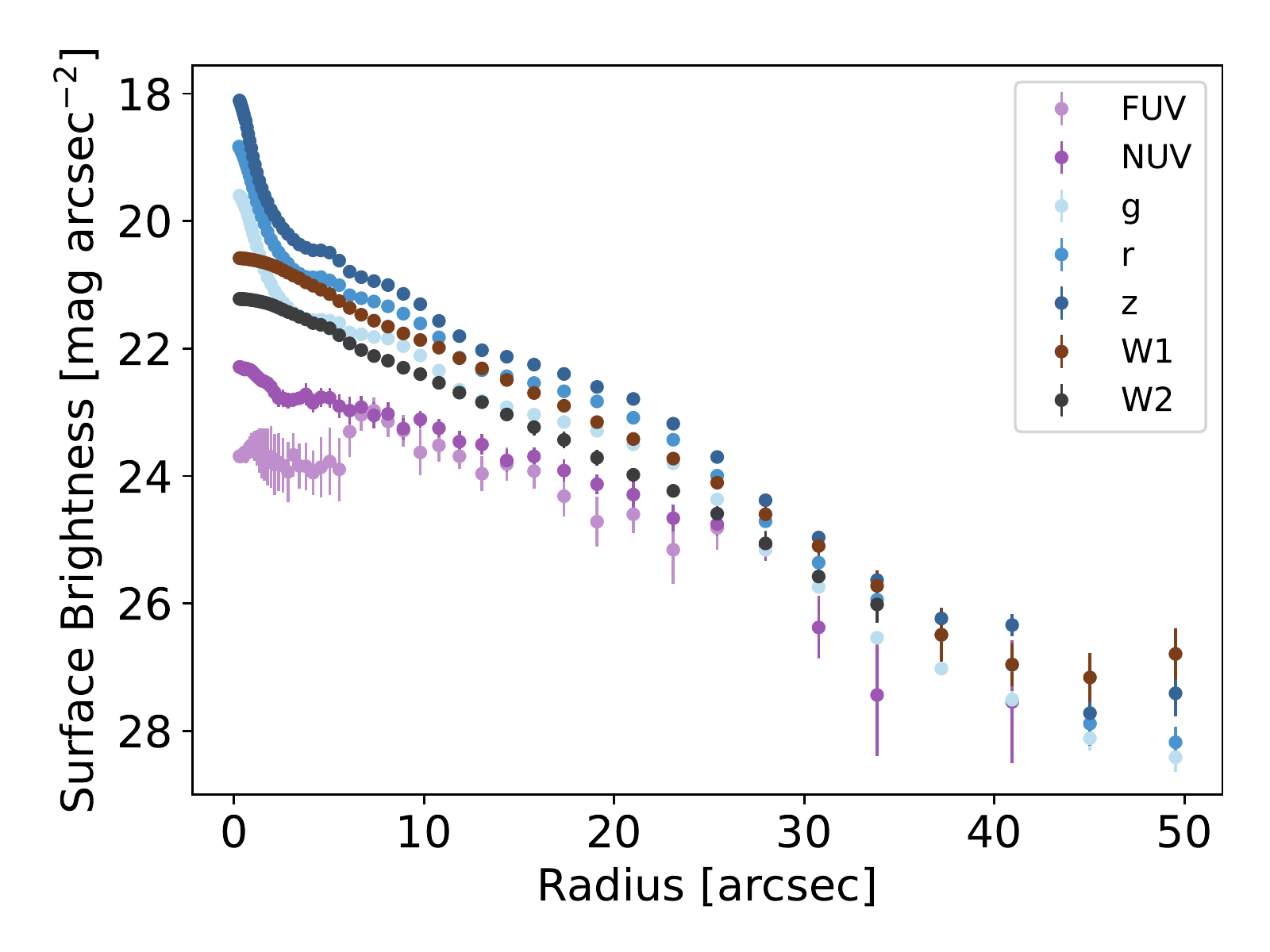}
    \caption{Truncated SB profiles of all seven photometric bands for ESO143-G028.
    Thanks to the forced photometry, each profile is evaluated using exactly the same isophotes, though each profile achieves different depths.
    Error bars are multiplied by 2 for visibility.}
    \label{fig:multibandprofile}
\end{figure}

\Fig{multibandprofile} shows the truncation results for each band, as well as the relative flux characteristics between different bands.
{\it FUV} and {\it NUV} have by far the lowest signal-to-noise, and low flux near the galaxy center which can be caused by heavy dust extinction and/or reduced star formation in the galaxy center.
DESI-LIS {\it g, r, z} bands have the largest dynamic range and resolution.
The {\it r}-band isophotal solution is also our reference point for the imaging of all the other bands.
The {\it W1, W2} bands have the lowest resolution, as can be seen by the absence  of structure in the central regions of ESO 143-G028 (compared to {\it g, r, z} bands), though the signal-to-noise for these observations is high enough for reliable global measurements at large radii.

\subsubsection{Photometric Corrections}\label{sec:photometriccorrections}

Before computing structural parameters, the truncated PROBES photometry (see \Sec{profiletruncation}) was corrected for Galactic extinction, K correction, and cosmological dimming.
The form of the global photometric correction is given as:

\begin{equation}
    A_c = A_o - A_g - A_K - A_z,
\end{equation}

where $A_c$ is the corrected brightness value, $A_o$ is the observed value, $A_g$ is the Galactic extinction, $A_K$ is the K correction, and $A_z$ is a cosmological dimming factor.
We do not add any uncertainty to our model from these corrections, though there is certainly a systematic uncertainty introduced with them.
We take Galactic extinctions for the DESI-LIS {\it g, r}, and {\it z} bands from \citet{Schlafly2011}; for the GALEX {\it FUV} and {\it NUV} bands, we use $E(B-V)$ values from \hbox{\citet{Burstein1982}} and transformations from \citet{Peek2013};
for unWISE {\it W1} and {\it W2} fluxes, we assume no extinction and therefore apply no correction.
K corrections were applied to {\it FUV, NUV, g, r,} and {\it z} bands using the calculator of \mbox{\citet{Chilingarian2012}}\footnote{See \url{http://kcor.sai.msu.ru/}}; for {\it W1, W2} bands we use the corrections from \citet{Jarrett2017}. 
K corrections for the PROBES sample are small given the low redshift of the sample, typically less than \wunits{0.1}{mag} for {\it FUV, NUV, g, r, z} bands and less than \wunits{0.5}{mag} for {\it W1, W2} bands.
Our cosmological dimming correction takes the form:
\begin{equation}
    A_z = 2.5\log_{10}((1+{\rm z}_{\rm helio})^3),
\end{equation}
where z$_{\rm helio}$ is the heliocentric redshift.
The SB scales as $(1+{\rm z}_{\rm helio})^3$ for the measurement of specific intensities (flux per Hz per steradian; e.g. in the AB magnitude system), in contrast with $(1+{\rm z}_{\rm helio})^4$ for bolometric fluxes~\citep{Tolman1930,Stabenau2008,Whitney2020}.
This correction is small for the relatively nearby PROBES galaxies, typically ranging from \wunits{0.005-0.05}{mag} over the range of our data.

\subsubsection{Photometry Inclination Correction}
\label{sec:inclinationcorrection}

Our inclination correction follows a similar procedure to \citet{Stone2021a} which corrects parameters to face-on equivalent, however with a slight variation. 
In \citet{Stone2021a}, the procedure to correct structural parameters was applied after they were extracted; here, we first correct SB profiles (individual isophotes) at all radii before computing structural parameters.
Therefore, all derived quantities are assumed inclination corrected as a consequence.
The functional form of the inclination correction to surface brightnesses is:

\begin{eqnarray}
    \mu_{\lambda}^c(R) = \mu_{\lambda}(R) + \gamma_{\lambda}\log_{10}(\cos(i_{\rm last})),  
\end{eqnarray}

where $\mu_{\lambda}^c$ is the corrected surface brightness, $\lambda$ is the band being corrected, $\mu_{\lambda}$ is the observed surface brightness, $i_{\rm last}$ is the disk inclination taken at the last isophote in the {\it r}-band profile, and $\gamma_{\lambda}$ is the correction coefficient.
The last isophote is used to determine the inclination as this most directly samples the disk, which gives a nominal representation of disk inclination and avoids other features such as bars and bulges.
AutoProf only fits ellipticity and PA above a S/N threshold (S/N per pixel $>3$ on average); beyond that point, AutoProf fixes these values and our inclinations are effectively measured at $\sim 25.5 \pm 1.0$ \magss in the {\it r}-band.
Since we use forced photometry from the {\it r}-band to all others, the inclination is the same regardless of band.

To determine the correction coefficients, we have performed least squares fits between $-\log_{10}(\cos(i))$ and the $\lambda - W1$ colour for all SB profile measurements in the PROBES dataset.
We use the {\it W1}-band as our reference point since it is minimally affected by dust and has higher S/N than the {\it W2}-band.
Since {\it W1} is assumed stable with inclination, we can use any dependence on $\lambda - W1$ colour to correct $\lambda$.
SBs fainter than \wunits{26}{\magss}, and inclinations greater than \wunits{80}{deg}, were not included in the calculation of corrections. 
Note that we only directly correct the SB profiles, curves of growth are recomputed by integrating the corrected SB profile.

\begin{figure*}
    \centering
    \includegraphics[width=\textwidth]{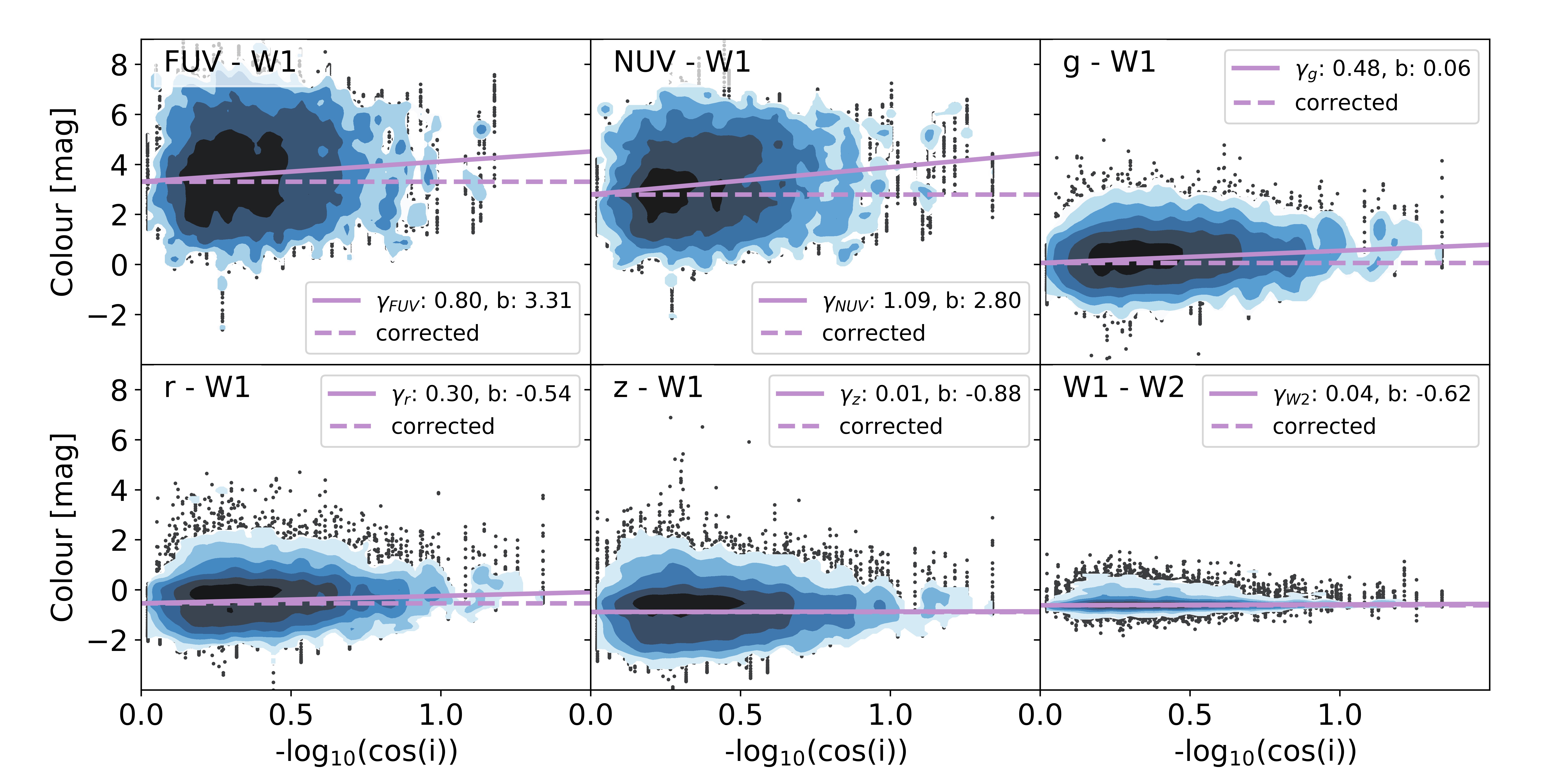}
    \caption{Linear least-squares fits for the inclination correction schemes applied to the PROBES magnitudes. 
    Every point in the corresponding PROBES photometry sample SB profiles is represented in the scatter plot. 
    Fits only include data points beyond $R_{\rm e}$. 
    The least-squares fitted slopes give the correction factor $\gamma_{\lambda}$, and the dashed line shows the corrected trend. 
    The bottom right figure shows that the $W2-W1$ colour is essentially independent of inclination.}
    \label{fig:inclinationcorrections}
\end{figure*}

\Fig{inclinationcorrections} shows the fitted inclination correction schemes against the raw PROBES surface brightnesses.
As expected, the redder bands are found to require less correction. 
As a check, the same procedure applied to $W1 - W2$ colours gives essentially no correction (just like the {\it z}-band). 
This straightforward correction scheme is adequate for the general data release of the PROBES compendium; more sophisticated prescriptions are beyond the scope of this paper.

\subsection{Available Parameters}\label{sec:availableparameters}

\begin{figure*}
    \centering
    \includegraphics[width=0.85\textwidth]{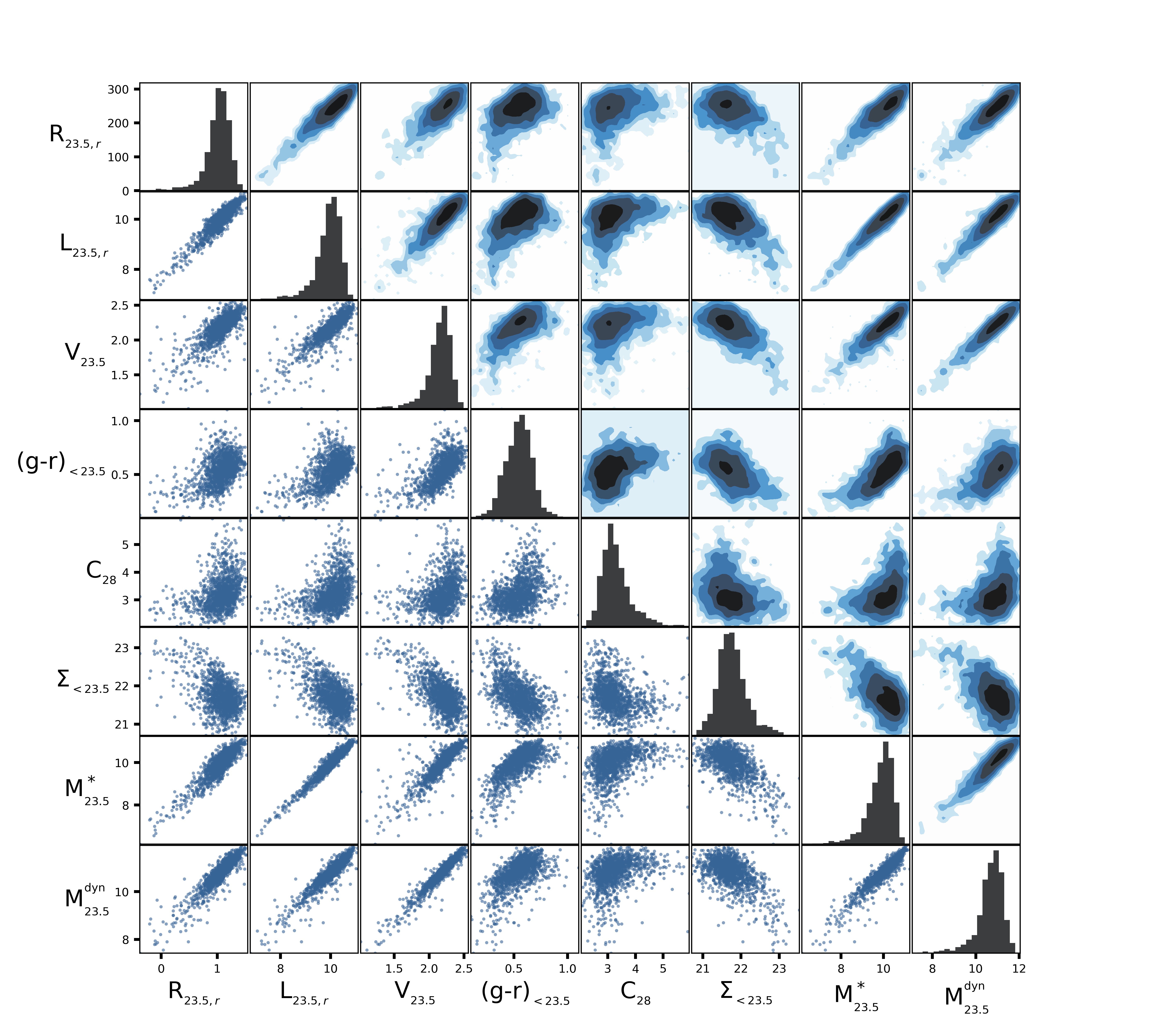}
    \caption{Triangle plot giving the distribution of a selection of structural parameters for the PROBES photometry sample. 
    (Lower triangle) The intersection of each row and column is a scatter plot of the relation between the two corresponding parameters; all parameters are shown in log scale. 
    The diagonal gives a histogram for each parameter. 
    (Upper triangle) The density plots show the distribution where scatter points overlap in the lower triangle. 
    Velocities are determined with the \citetalias{Courteau1997} model.}
    \label{fig:triangleplot}
\end{figure*}

We provide a vast array of structural parameters for the PROBES photometric sample, covering a wide range of radii for all galaxies.
A selection of parameters, evaluated at $R_{23.5}$, are presented in \Fig{triangleplot} showing the scope of scaling relations available with the data at almost any desired radius.

\subsubsection{Radii}\label{sec:evalradii}

\begin{figure}
    \centering
    \includegraphics[width=\columnwidth]{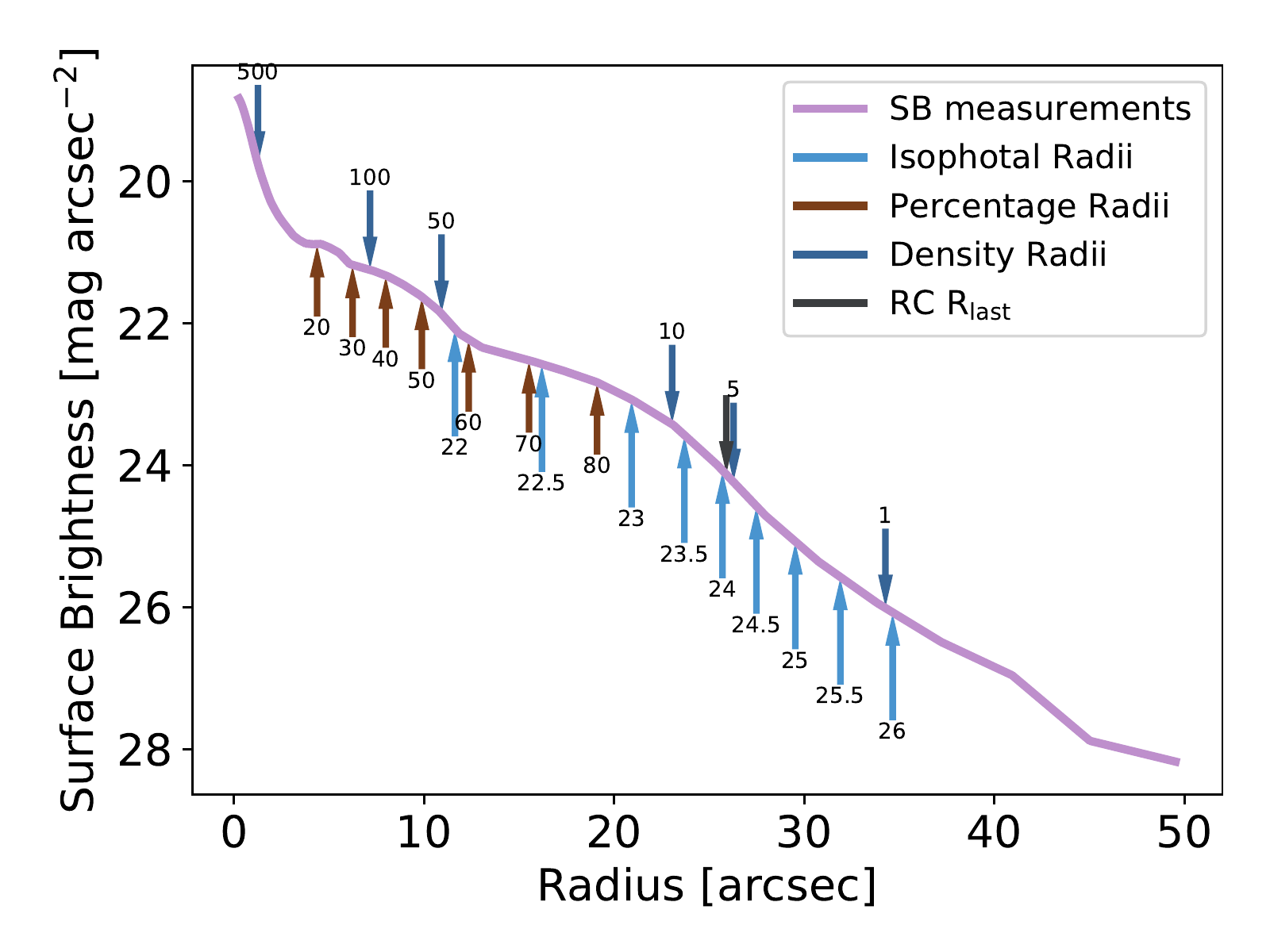}
    \caption{Example {\it r}-band SB profile for \emph{ESO143-G028} with various radial metrics indicated with arrows. 
    For instance, isophotal radii correspond to SB levels ranging from 22 to \wunits{26}{\magss} in steps of \wunits{0.5}{\magss}; see text for details.
    These radii are evaluated for all galaxies in the PROBES photometry sample, most other structural parameters are evaluated at all of these radii. 
    Errors on the SB profiles are not shown as they are too small.}
    \label{fig:radiusmeasures}
\end{figure}

It is useful to evaluate a number of fiducial apparent radii for the measurement of other structural parameters.
We have adopted the {\it r}-band SB profiles to calculate our fiducial radii, as these profiles have high signal-to-noise, good spatial resolution, and low dust extinction.
It is also for these reasons that our photometry uses {\it r}-band profiles for forced photometry at other bands.
For isophotal radii, values from $R_{22}$ to $R_{26}$ in increments of \wunits{0.5}{\magss} are obtained as shown on \Fig{radiusmeasures}. 
Sampling brighter than 22~\magss would eliminate many galaxies, and reaching fainter than 26~\magss might expose interlopers. 
Light percentage radii are also computed from \wunits{20}{\%} up to \wunits{80}{\%} in increments of \wunits{10}{\%}, the most common of these being the effective radius $R_{50} = R_{\rm e}$. 
We also compute stellar mass percentage radii, though they are not shown in \Fig{radiusmeasures} since they are similar to light percentage radii; stellar mass percentage radii are better matched for comparisons with simulations.
Stellar surface density radii at \wunits{500,100,50,10,5,1}{$M_{\odot} {\rm pc}^{-2}$} also provide useful physical reference points in each galaxy.
$R_{\rm last}$, the radius of the last point in a galaxy's RC is also provided.
Ignoring integration times and instrument sensitivities, $R_{\rm last}$ can roughly be viewed as a tracer of the \Ha endpoint in a galaxy.
Whenever possible, parameters such as luminosity and stellar mass are also computed to infinity by extrapolating the outer SB profile as an exponential disk (see \Sec{evalluminosity}). 

\subsubsection{Velocity}\label{sec:evalvelocity}

Velocities are a critical and integral part of the PROBES compendium and we have extracted representative velocities at all fiducial radii discussed in \Sec{evalradii} using the \textit{tanh} and \citetalias{Courteau1997} RC models (\Sec{rotationcurves}). 

The \citetalias{Courteau1997} model captures the RC shapes well enough that reliable extrapolations can be performed for most of the PROBES sample.
As such, velocities are provided for all fiducial radii (\Sec{evalradii}) even if this involves an extrapolation of the RC.
It is straightforward for users to select interpolation measurements only, as desired. 
The two RC parameterizations have the form:

\begin{eqnarray}
        V_{\tanh}(r) &= V_{\rm sys} + V_{\rm max}\tanh((r - r_0)/r_t) \\
        V_{\rm C97}(r) &= V_{\rm sys} + V_{\rm max}\frac{(1 + r_t/(r - r_0))^{\beta}}{(1 + (r_t/(r - r_0))^{\gamma})^{1/\gamma}}
\end{eqnarray}

where $r$ is the radial location in the RC, $V_{\rm sys}$ is the systemic velocity of the 
galaxy, $V_{\rm max}$ is the maximum velocity\footnote{This is not strictly true for the \citetalias{Courteau1997} model, though it is a good proxy.}, $r_0$ is the center point offset, $r_t$ is the turnover radius from the rising to the flat regime, $\beta$ controls rising/falling strength of the RC outskirts, and $\gamma$ determines the shape of the RC turnover to ``flat.''

\begin{figure}
    \centering
    \includegraphics[width=\columnwidth]{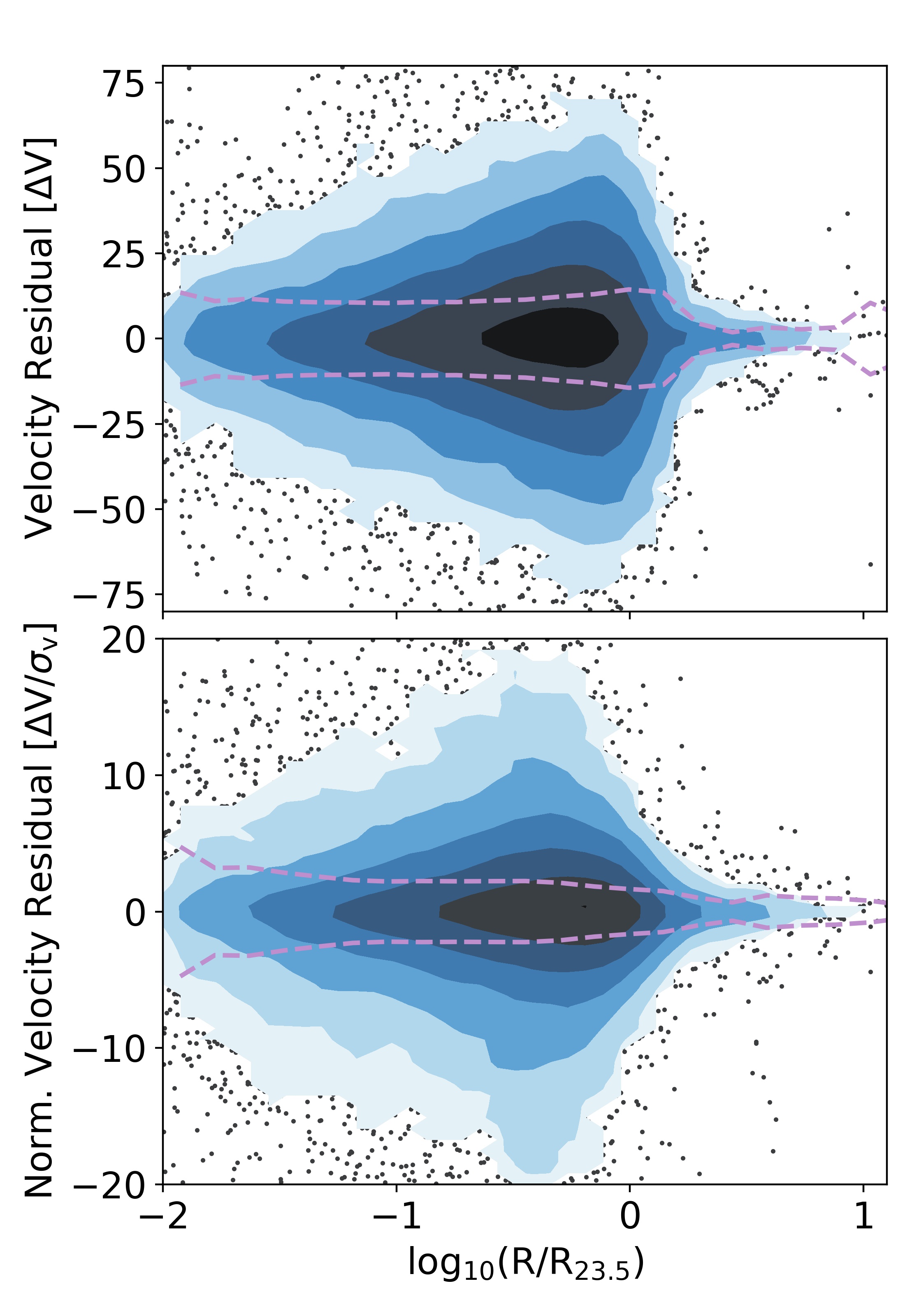}
    \caption{Distribution of velocity residuals (top), normalized by the error of the velocity measurement (bottom), for the \citetalias{Courteau1997} model fit to each RC in the PROBES photometry sample. 
    Radii are normalized by the isophotal radius, $R_{23.5}$. 
    The purple dashed lines give the \wunits{68}{\%} band of data.
    The axes limits are truncated to show the main trends.}
    \label{fig:rcresiduals}
\end{figure}

\Fig{rcresiduals} shows the \citetalias{Courteau1997} model residuals for all of the PROBES photometry sample.
The radii are normalized by $R_{23.5}$ to ensure that like regions between galaxies are being compared.
A number of interesting trends stand out in this figure.
The top figure shows that typical velocity residuals are within \wunits{15}{km s$^{-1}$}. 
The bottom figure makes clear that this offset cannot be explained solely by statistical errors since the typical errors exceed $1\sigma$. 
One must invoke noncircular motions, PSF smoothing, and other complex behaviours that are captured in the PROBES RCs.
A notable drop in residuals beyond $\sim R_{23.5}$ is partly due to the decline in the number of galaxies with extended RCs beyond that point and to \HI RCs being smoother than \Ha.
Most extended RCs that sample beyond $R_{23.5}$ come from the SPARC survey (see \Sec{sparc}) which uses \HI as a velocity tracer.

The absolute velocity residuals (\Fig{rcresiduals} upper panel) are more or less constant within $R_{23.5}$, though the contours show that most velocity measurements occur prior to $R_{23.5}$.
The relative velocity residuals (\Fig{rcresiduals} lower panel) increase considerably towards the center, in the central regions (within $R_{23.5}$) $\Delta V/\sigma_v \approx 2.5$ when in principle it should be $1$.
This is expected with the central regions being more dominated by noncircular motions or other complex behaviours which are not modelled or accounted for in the error values.
The large relative central errors may also be caused by PSF blur which can affect velocity measurements especially at the center.
The outer regions (beyond $R_{23.5}$) do have $\Delta V/\sigma_v \approx 1$ as expected for well behaved large disks.
For individual cases, it is likely that noncircular motions exist at large radii as well~\citep{Spekkens2007,Widrow2014,Nandakumar2022}; however, they are not apparent in aggregate.

We also performed the same residual analysis using \emph{tanh} model fits instead of the \citetalias{Courteau1997} model.
The resulting distributions were nearly identical, though the average normalized residual in the central regions was $\approx 3.7$ (instead of $2.5$) owing to the lower expressive power of \emph{tanh} models to fit more complex features present in the PROBES RCs (e.g. rising/falling outer profiles or bulge feature).

\subsubsection{Luminosity}\label{sec:evalluminosity}

The apparent magnitude, absolute magnitude, and luminosity corresponding to each radius specified 
in \Sec{evalradii} are computed in all {\it FUV, NUV, g, r, z, W1, W2} bands. 
These are computed directly from the AutoProf curve of growth.

Using a linear interpolation of the curve of growth, the apparent magnitude is computed and then transformed into the absolute magnitude and luminosity according to:

\begin{eqnarray}
        & M_{\lambda} &= m_{\lambda} - 5\log_{10}(D/10)  \\
        & L_{\lambda} &= 10^{(M_{\odot,\lambda} - M_{\lambda})/2.5}, 
\end{eqnarray}

where $M_{\lambda}$ is the absolute magnitude in the $\lambda$ band, $m_{\lambda}$ is the apparent magnitude, $D$ is the distance in parsecs, $L_{\lambda}$ is the luminosity, and $M_{\odot,\lambda}$ is the corresponding magnitude of the sun.
We take our $M_{\odot}$ values from \citet{Willmer2018} for the {\it FUV, NUV, g, r, z, W1, W2} bands, giving $17.30$, $10.16$, $5.05, 4.61, 4.50, 5.92,$ and $6.58$ respectively.

The extrapolation to infinity involves fitting an exponential disk through all SB points beyond $R_{60}$, the radius which encloses $60\%$ of all the detected light.
The contribution of the exponential disk beyond the last point in the SB profile is added to the total magnitude and further calculations are performed accordingly.
The choice of $R_{60}$ as the reference point comes from manual experimentation, though other choices do not alter the resulting total magnitudes appreciably (see Section 4.3 of \citealt{Courteau1996}). 

\subsubsection{Colours}\label{sec:evalcolours}

Colours are a useful tracer of stellar populations in a given galaxy \citep{Conroy2013,Courteau2014}.
These may be evaluated at or within a given radius, and both modes are included in our tables.
Colours within a given radius are determined by taking a difference in apparent magnitudes interpolated from a curve of growth, whereas colours at a given radius are determined by taking the difference in SB values.

\subsubsection{Concentration}\label{sec:evalconcentration}

Concentration is a useful morphological metric for the quantitative comparison of the structure of light profiles \citep{Conselice2014}.
Various definitions of concentration exist, for a nonparametric concentration we use the following definition:  

\begin{eqnarray}
    C_{28} &= 5\log_{10}(R_{80}/R_{20})\label{equ:c28}
\end{eqnarray}

which measures the ratio of the radii enclosing 80 and 20 percent of the total light of a galaxy.
As a nonparametric measure, $C_{28}$ is universally applicable since it applies to any light profile.

The PROBES compendium tables also include the S{\'e}rsic index \citep{sersic1968} as a parametric concentration parameter.
The form of the S{\'e}rsic function for fitting galaxy light profiles is given below, 

\begin{eqnarray}
    I(R) &= I_e{\rm exp}\left(-b_n\left[\left(\frac{R}{R_e}\right)^{1/n}-1\right]\right),\label{equ:sersic}
\end{eqnarray}

where $I(R)$ is the intensity as a function of projected radius, $R$, $I_e$ is the intensity at the effective radius, $b_n$ is a normalization constant, $R_e$ is the effective projected radius, and $n$ is the S{\'e}rsic index.
Because of the straightforward interpretation of its model parameters, the S{\'e}rsic function has been widely used. 
However, the reliable reproduction of S{\'e}rsic parameters between studies can be especially challenging~\citep{Arora2021,Sonnenfeld2022}.

\subsubsection{Surface Brightness}\label{sec:evalsurfacedensity}

Surface brightness (SB), in units of mag arcsec$^{-2}$ in any given photometric band, is a distance-independent quantity (for noncosmological applications) that informs us about the density of galactic systems. 
A common measure of galaxy brightness is the central SB, which simply takes the SB at $R=0$ from a SB profile.
The latter is however especially sensitive to seeing variations between observatory sites, between observations at the same site, and dust extinction for bluer bands.  
The mean SB, $\Sigma_{\rm in}(<R)$, measured within various characteristic radii, which can extend beyond a few seeing lengths, is more robust to seeing fluctuations.
It is defined as: 

\begin{equation}
    \Sigma_{\rm in}(<R) = m(R) + 2.5\log_{10}(\pi qR^2),
\end{equation}

where $m(R)$ is the apparent magnitude within a galactocentric radius $R$ and $q$ is the axis ratio.

One may also calculate the SB at a given radius.
The local SB at each radius is computed via interpolation of the SB profile.
Note that we still report these values for isophotal radii despite their definition being connected to SB values.
Our isophotal radii are determined in the {\it r}-band and will typically have different SB values in other photometric bands; i.e., the {\it r}-band $R_{25}$ will not have an SB of \wunits{25}{\magss} in the {\it g}-band.

\subsubsection{Stellar Mass}\label{sec:evalstellarmass}

The inferred stellar mass within a given radius, or $M^*(R)$, is one of the most fundamental galaxy structural parameters as it enables a direct comparison with theoretical predictions. 
Unfortunately, stellar masses also carry a large uncertainty in light of the many assumptions that their computation entails \citep{Courteau2014}.
Most photometry-based assessments of stellar mass rely on broadband colour transformations, which themselves make assumptions about the stellar populations makeup of a galaxy. 
The PROBES photometry sample relies on five broadband fluxes in order to constrain the distribution of stellar populations by mass in galaxies\footnote{GALEX measurements are left out since they do not enter the colour-mass-to-light transformations of \citet{Cluver2014} and \citet{Roediger2015} used in this paper.}. 
We have used the colour-mass-to-light transformations of \citet{Cluver2014} and \citet{Roediger2015} to infer stellar mass estimates.
For each stellar mass estimate, a colour and a luminosity are needed from a given band.  
Our final value of $M^*(R)$ is the average of the four resulting stellar mass estimates from the following colour combinations: ({\it g-z, g}); ({\it g-r, r}); ({\it r-z, z}); and ({\it W1-W2, W1}).
The standard deviation of this estimate gives the reported error.

In general, systematic errors on stellar mass estimates exceed the random errors~\citep{Roediger2015}; by reporting the scatter in the $M^*$ estimates from four band combinations, we present a hybrid of random and systematic errors. 

\subsubsection{Stellar Surface Density}\label{sec:evalstellarsurfacedensity}

The combination of \Secs{evalsurfacedensity}{evalstellarmass} enables us to compute a stellar surface density, $\Sigma^*$, in units of ${\rm M}_{\odot}{\rm pc}^{-2}$. 

\begin{equation}
    \Sigma^* = \frac{M^*}{L}10^{-(M_{\odot} - \Sigma_{\rm in})/2.5} .
\end{equation}

As stellar mass should be band independent, we averaged multiple bands for the computation of $\Sigma^*$ following the same procedure as \Sec{evalstellarmass}.

Stellar surface density profiles allow one to evaluate radii at which certain densities are reached \citep[e.g.][]{Trujillo2020}.
Such radii are included in our table of structural parameters for reference (\Sec{datatables}). 

\subsubsection{Dynamical Mass}\label{sec:evaldynamicalmass}

Required for studies of the dark matter distribution and gravitation potentials in galaxies, the dynamical mass gives an estimate of the total enclosed mass (baryons+dark matter) of a galaxy \citep{Courteau2014}. 
A significant product of the PROBES sample is indeed the deep long slit RCs available for every galaxy.
These give reliable estimates, within $\sim$\wunits{10}{\%}~\citep[Sec.~2.6, Fig.~2.17]{Binney2008} of the total enclosed mass within a galactocentric radius $R$. 
Here we opt for a straightforward mass calculation using a spherical mass distribution:

\begin{equation}
    M_{\rm dyn}(R) = \frac{V^2(R)R}{G},
\end{equation}

where $M_{\rm dyn}(R)$ is the estimate of dynamical mass enclosed within radius $R$, $V(R)$ is the circular velocity at radius $R$, and $G$ is the gravitational constant.
This mass estimate differs slightly for nonspherical mass distributions \citep{Binney2008}, however it is suitable for most applications and does not require specifying an arbitrary or ill-defined mass distribution.

\subsection{Parameter Uncertainties}\label{sec:uncertainties}

All structural parameters have an associated standard error.
This is computed by propagating uncertainty from measurement errors using the standard first-order method unless otherwise specified,

\begin{equation}
    \sigma_f^2 = \sum_p\left(\sigma_p\frac{df}{dp}\right)^2,
\end{equation}

where $\sigma_f$ is the error on some parameter with functional form $f$, and $p$ is a parameter which is used in the calculation of $f$.
These errors do not include covariances between other parameters, as such they are not suitable for certain Bayesian error propagations.

\section{Data Tables}\label{sec:datatables}

The raw PROBES data are composed primarily of RCs and SB profiles. 
The RCs are formatted as shown in \Tab{rotationcurves} in a csv file.
These data are presented mostly as in the original survey, except with some concessions for homogenization.
Velocities are left uncorrected for inclination and redshift.
In all cases the systemic velocity of the galaxy has been removed and the RCs have been recentered (see \Sec{evalvelocity}).
All profiles have been flipped such that the approaching side has positive radius values and the receding side has negative radius values.
In cases where multiple observations exist, all available RCs were stacked after centering.
No averaging was performed, the final combined profile for a given galaxy simply includes all data points from all available profiles for that galaxy.
In some cases, RC uncertainties were not provided and we used as a proxy the standard deviation of the residuals to a \citetalias{Courteau1997} model fit; these errors per point typically range from \wunits{5-15}{km s$^{-1}$}.
Using the standard deviation of the residuals typically gives conservatives errors by assuming all deviations from the model are due to measurement errors when in fact some may be due to noncircular motions.

The SB profiles are formatted as shown in \Tab{sbprofcolumns} following the format in \citet{Stone2021b}.
These profiles are outputted by AutoProf; fitting is performed on the {\it r}-band profiles and forced photometry, which ensures that colours can be evaluated on the same pixels, is applied for the other bands.
For the fitting procedure, we ran the standard AutoProf pipeline and also extracted (wedge) radial profiles. 
AutoProf performed a number of checks to ensure that most fits converged to reasonable solutions.
In some cases, we used fixed centers and/or masks to improve the fits.
Still, a fraction of galaxies, \wunits{15.7}{\%}, failed at least one check.
These failures may result from slight asymmetries, oversaturated pixels, or nearby bright sources.
Typically, if only a single flag is raised then the photometry is still adequate for most purposes.
Only \wunits{2.5}{\%} of galaxies have two or more AutoProf flags; for most analyses, these galaxies should be discarded.
They are however kept in the sample as users may wish specifically to look at these systems for their complexity.
However, in such cases the users should manually examine the photometry and consider using the (wedge) radial profiles along the major/minor axes instead of the full ellipse solution.

\begin{table}[h]
    \centering
    \caption{Rotation curve columns}
    \begin{tabular}{ c  c  l }
    \hline
        Column & Units & Description \\ 
        (1) & (2) & (3) \\
        \hline
        R & arcsec & Semi-major axis location \\
        V & km s$^{-1}$ & Circular velocity (\Sec{evalvelocity}) \\
        V\_e & km s$^{-1}$ & Circular velocity measurement error \\
        \hline
    \end{tabular}
    \label{tab:rotationcurves}
\end{table}

\begin{table*}
\caption{Surface brightness profile columns}
\label{tab:sbprofcolumns}
\begin{tabular}{ccl}
\hline 
Column & Units & Description \\ 
(1) & (2) & (3)\\
\hline 
R & arcsec & Isophote semi-major axis length ($=a$) \\
SB & \magss & Median SB along isophote \\
SB\_e & \magss & Uncertainty on SB estimate \\
totmag & mag & Total magnitude enclosed within isophote, computed by integrating SB profile \\
totmag\_e & mag & Uncertainty in totmag estimate propagated through integral \\
ellip & \nodata & Ellipticity of isophote ($=1-b/a$, where $b=$ isophote semi-minor axis length) \\
ellip\_e & \nodata & Uncertainty in ellipticity estimate determined by local variability \\
pa & deg & Position angle of isophote relative to positive $y$-axis (increasing counter-clockwise) on the image \\
pa\_e & deg & Uncertainty in PA estimate determined by local variability \\
pixels & count & Number of unmasked pixels sampled along isophote or within band \\
maskedpixels & count & Number of masked pixels rejected along isophote or within band \\
totmag\_direct & mag & Total magnitude enclosed in current isophote by direct pixel flux summation \\
\hline
\end{tabular}
\end{table*}

Along with the raw PROBES data (RCs and SB profiles), three tables are provided with information about the PROBES galaxies.
The ``main\_table.csv'' file includes basic data about all PROBES compendium galaxies: 
position (RA,DEC), redshift, distance, and photometry information.
The latter indicates whether galaxies have photometry and if they were flagged by AutoProf processing.

\begin{figure}
    \centering
    \includegraphics[width=\columnwidth]{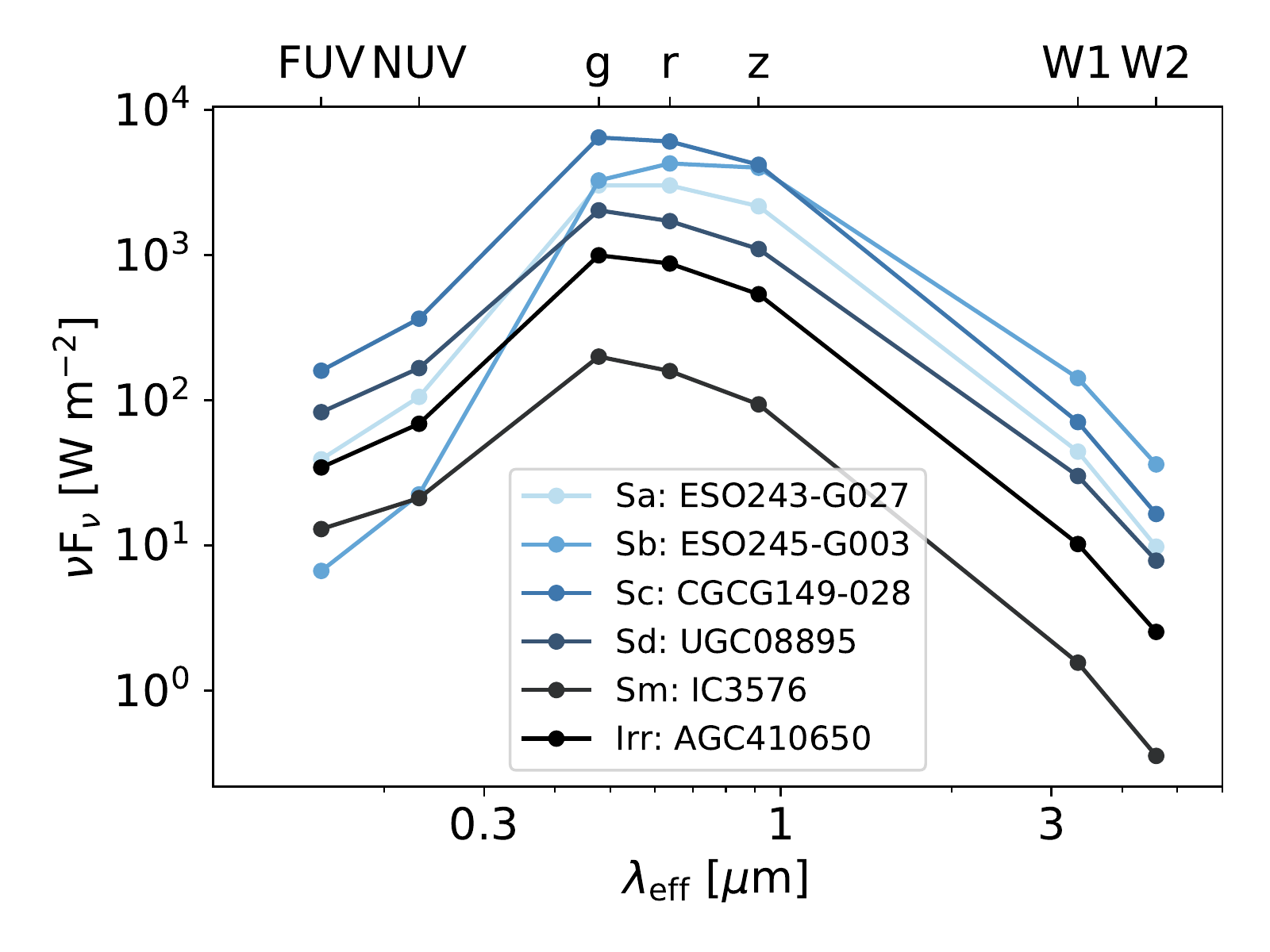}
    \caption{Typical spectral energy distributions taken from the PROBES structural parameters table for a representative range of galaxy morphologies. }
    \label{fig:sedexample}
\end{figure}

The ``structural\_parameters.csv'' file includes the bulk of the structural parameters information for the PROBES photometry sample.
The number of available parameters is very large and cannot easily be recounted here.
A consistent formatting scheme is meant to offer an unambiguous interpretation of each column.
A given parameter is divided into two parts by a ``$|$'' symbol with the left side indicating the parameter being computed and the right side indicating the radial location of that measurement (or any other necessary information about computing that parameter).
For example, the column ``Col\_in:g:r$|$Ri23.5:r'' represents the integrated $g-r$ colour within the isophotal radius $R_{23.5}$ measured in the {\it r}-band.
\Fig{sedexample} shows a typical spectral energy distribution constructed from the structural parameters table.
Absolute magnitudes were taken from columns ``absMag:$\lambda|$Ri26:r'' where $\lambda$ is the filter; these were converted to flux units for the figure. 

Further, ``Mstar$|$Rp80:r'' represents the stellar mass enclosed within the $80\%$ total light radius as determined from the {\it r} band profile.
This formatting convention is held for all parameters.
When stellar mass is used instead of a wavelength band, the ``*'' character takes its place.
Finally, the ``model\_fits.csv'' file gives the parameters for standard fitted models.
The fitting parameters for the \textit{tanh} and \citetalias{Courteau1997} models applied to the rotation curves are reported here, along with S{\'e}rsic fits to each band-specific SB profile.
 
\section{Conclusion}\label{sec:conclusion}

We have presented the PROBES compilation of deep RCs for \probesnedunique galaxies and matching multiband photometry for \probesphotometric of them.
An overview of the data characteristics and the convenient formatting for general use was also discussed.
The raw data, as well as a large number of homogeneously determined structural parameters, are available for download within the supplementary material.

By linking deep kinematics and photometry for a statistically significant sample, PROBES offers a comprehensive picture of late-type galaxy structure.
Updates to the PROBES data base are anticipated as new data become available.
For instance, Frosst et al (PROBES-II, in preparation) will emphasize low mass dwarf systems with the addition of 716 nearby galaxies with high-quality rotation curves and 578 matching surface photometry profiles derived from DESI-LIS photometry. 

\begin{acknowledgments}
We are grateful to the Natural Sciences and Engineering Research Council of Canada, the Ontario Government, and Queen's University for support through various scholarships and grants.
Special thanks go to Arjun Dey and John Moustakas for discussions about the DESI Legacy Imaging Survey.
This research has made use of the NASA/IPAC Extragalactic Database (NED),
which is operated by the Jet Propulsion Laboratory, California Institute of Technology, under contract with the National Aeronautics and Space Administration.

\end{acknowledgments}

\vspace{5mm}

\software{astropy~\citep{astropy},  
AutoProf~\citep{Stone2021b}
          }

\bibliography{PROBES}
\bibliographystyle{aasjournal}

\end{document}